\documentclass{cpeauth}

\usepackage{times}
\usepackage{graphics}
\usepackage{subfigure}
 \usepackage{lscape}
 \usepackage{multirow}
  \usepackage{parcolumns}
  \usepackage{caption}
  \usepackage{setspace}
  \usepackage{slashbox}
  \usepackage{wrapfig}

\usepackage{amsmath}
\usepackage{amssymb}
\usepackage{algorithm} % Algoritmos
\usepackage{algorithmic} % Algoritmos

\begin{document}

\title { Memory Aware Load Balance Strategy on a Parallel Branch-and-Bound Application}

\author{Juliana M. N. Silva\\
Cristina Boeres\\
L\'ucia M. A. Drummond\\
Artur A. Pessoa \\
 Univesity Federal Fluminense -  UFF \\ 
Rua Passo da Pátria 156 - Bloco E. São Domingos Niter\'oi - RJ
\\ jsilva@ic.uff.br\\ boeres@ic.uff.br\\ lucia@ic.uff.br\\
% \and
% First line of institution2 address\\ Second line of institution2 address\\ 
% SecondAuthor@institution2.com\\
}

\thispagestyle{empty}

\begin{abstract}
The latest trends in high-performance computing systems show an increasing demand on the use of a large scale multicore systems in a efficient way, so that high compute-intensive applications can be executed reasonably well. However, the exploitation of the degree of parallelism available at each multicore component can be limited by the poor utilization of the memory hierarchy available. Actually, the multicore architecture introduces some distinct features that are already observed in shared memory and distributed environments. One example is that subsets of cores can share different subsets of memory. In order to achieve high performance it is imperative that a careful allocation scheme of an application is carried out on the available cores, based on a scheduling model that considers the main performance bottlenecks, as for example, memory contention.
In this paper, the {\em Multicore Cluster Model} (MCM) is proposed, which captures the most relevant performance characteristics in multicores systems such as the influence of memory hierarchy and contention. Better performance was achieved when a load balance strategy for a Branch-and-Bound application applied to the Partitioning Sets Problem is based on MCM, showing its efficiency and applicability to modern systems.
\end{abstract}
\maketitle

%------------------------------------------------------------------------- 
\section{Introduction}

Multicore architectures have become dominant today due to the considerable enhancement on computing systems performance. Multicores can be found in a variety of domains. Currently, high performance platforms like clusters are composed of multicore nodes or multicore clusters connected by network channels. These modern platforms suggest a hierarchical memory: cores that belong to the same processor can share caches, cores belonging to different processors share main memory (like RAM or DRAM) and cores that belong to different nodes do not share any memory resource \cite{Savage, Song2011}. 

Parallel applications could benefit from such memory hierarchy to improve performance. The use of cache as shared memory can reduce the communication time between the tasks of an application, and, therefore, tasks that communicate more frequently should be placed in cores that share cache, avoiding communications in main memory or message passing over the network \cite{Song2011, Sadaf, Song}. However, depending on the amount of memory required for communicating and computing tasks, allocating tasks in many cores that are sharing the cache may exceed its capacity, making necessary too many accesses to main memory. These accesses can cause a bottleneck in the channels and worsen the application performance \cite{Savage, Song, Mars, Rashid}. 

Using the environment characteristics in order to improve application performance is not new. For doing so, it is necessary to define models that represent the most relevant features of the environment where the application will run. Nonetheless, this is not an easy task and scheduling algorithm or load balance strategies should be based on such a model  and providem better application's runtime.

This paper proposes the {\em Multicore Cluster Model} (MCM), which was based on an extensive set of experiments of a synthetic application that identifies the potential bottlenecks promoted by sharing memory resources and their impact when executing computation and communication tasks. The model considers three levels of communication: i) the communication made through shared memory by intra-chip cache, ii) through inter-chip shared memory and iii) communication between cluster nodes via messages. Scheduling and load balance strategies should be adjusted considering the architecture model and the characteristics of the application, so that it takes the maximum advantage  of the execution environment. A long these lines, a load balance strategy for a class of branch-and-bound application based on MCM is also proposed.

In order to evaluate and validate our proposals, a parallel branch-and-bound algorithm applied to the set partitioning problem ($PBB_{SPP}$) was developed based on a load balance mechanism also introduced here. The experiments confirm that the model represents relevant features of the architecture which affect the application performance. The results showed that when memory access bottlenecks are avoided, the execution time of $PBB_{SPP}$ can be improved by up to 70\%.  

Summarizing, the main contributions of this work are the following:

\begin{enumerate}

\item  A new  model that considers not only the relevant architectural characteristics of processing and communication via different levels of memory and network in a multicore cluster, but also how those characteristics are impacted by the amount of memory required by the application tasks. Thereby, the  impact that the quantity of memory required by processing and communicating tasks on the execution and communication costs where measured and modeled. The objective here is to provide a model that includes  into the typical processing and communication  costs, the one associated with contention in the different levels of memory.

\item  Based on the model, a  novel load balance strategy is proposed in which the memory hierarchy is accounted when communication is held and the quantity of data allocated to each task is evaluated so that the work load is balanced, avoiding therefore memory contention bottlenecks.

\item  Finally, a real application  based on the branch-and-bound algorithm was used to validate the proposed work.   In the related literature, there is a large number of papers about parallel branch-and-bound, but, to the best of our knowledge, few of them were designed to take advantage of a computing system with  both shared and distributed memory.   The implementation of the parallel branch-and-bound used here was based the proposed load balance strategy.

\end{enumerate}

The remaining of this paper is organized as follows. Section 2 presents the related literature about high performance architecture models. A set of tests used to identify the relevant characteristics of multicore clusters and a  new load balance  mechanism based on the obtained results are introduced in Section 3. Section 4 presents the use of the proposed load balance strategy  in a parallel branch-and-bound  to  solve the Set Partitioning Problem. Experimental results  and analysis, aiming to evaluate the efficiency of the resulting application, are shown in Section 5. Section 6 concludes the paper.
%------------------------------------------------------------------------- 

%------------------------------------------------------------------------- 

%------------------------------------------------------------------------- 
\section{High Performance Platforms Models}
\label{modelo_arquitetura}

Due to the variety of parallel and distributed architecture, it is difficult to define a precise and yet general model of parallel computation. On the attempt to identify the actual trend, this section outlines  models of parallel computation with the aim to identify the relevant characteristics that must be considered when executing parallel applications. 

It is already well stablished the distinction between distributed memory, where each processor has its own local memory, and shared memory, where all processors have access to a common memory. For many years, high performance computing was developed based on distributed systems mainly  due to their potential to solve much larger problems and their scalability. However, at the same time, in order to improve the performance of processors even further, architectural designers put together more and more processor cores on the same chip, promoting the multicore advent. In this case, good performance relies on the software ability to exploit the shared memory hierarchy. For doing so, it is important to define a computation model that incorporates the parameters of parallel architectures that are essential to characterize the  parallel systems.

\subsection{Model for shared memory architecture}

The Parallel Random Access Machine (PRAM)  model~\cite{Wyllie} consists of a number of processors, each of which computes one instruction in one time unit, on different data, synchronously, and then communicates via shared memory, also within one step~\cite{JaJa}.
The great acceptance of the PRAM model by the theoretical community has been due to its simplicity and universality and a large number of parallel algorithms based in it have been designed. While the PRAM model is an idealistic one, unfortunately it is not a realistic.  Nevertheless,
much research effort has been expended on the  attempt to incorporating critical parameters
of parallel systems, mainly the ones related to communication overhead ~\cite{Cole, Gibbons, Gibbons2, Gibbons3, Maggs, Vijaya}.

In early 90's, due to the continuous technological advances on memory bandwidth and latency, the use of shared memory was a reality. Since the program designer  wish to take full advantage of the memory system, it is necessary to consider the time to access not only the local main memory but also the other several levels of  memory.
Aggarwal \textit{et al} in \cite{Aggarwal} proposed the Hierarchical Memory Model (HMM) designed to capture the effect of memory hierarchy. HMM considers a  random  access memory machine  where  access to  memory location  $x$  requires $\lceil log x \rceil$ time  instead  of  the  typical  constant  access. An extension of HMM, the HMBT, was proposed in ~\cite{Aggarwal2} in which a block of consecutive locations can be copied in constant time after the initial latency access is paid. However, both models do not consider parallel machines. Thus, \cite{Ben} introduced extensions of the HMBT to model memory systems in which data transfers between memory levels may proceed concurrently. 

Already in \cite{Bowen1}, the Parallel Memory Hierarchy (PMH) models a computer as a tree of memory modules with processors in the leaves. The main characteristic is the representation of the transfer cost of a block of data between the tree nodes. In \cite{Bowen}, the Uniform Memory Hierarchy (UMH) is proposed, the cost of data movement between different levels  of  the memory hierarchy. Although the works above mentioned are  two decades old, it is interesting to note the  evidence of current architectures characteristics such as multicore clusters, especially the relative impact of the memory hierarchy in the performance of applications. These set of works however, lack mainly on modeling both distributed and shared memories.

Gibbons \textit{et al} in \cite{Gibbons2} introduced the Queuing Shared Memory (QSM) model, which accounts for limited communication bandwidth while still providing a simple shared-memory abstraction. The QSM model consists of processors with individual private memory as well as a global shared memory. However this model ignores the memory hierarchy in a processor.

\subsection{Model for distributed memory architecture}

With the objective of designing a scalable system, distributed memory networks have become the main stream for the specification of an efficient solution for very large dimension problems. However, the performance of these proposed solutions can be affected by the limitation on bandwidth and latency on communications. Many researchers have evaluated the behavior of distributed memory architectures, with the aim of designing a general
purpose parallel model. The Distributed Memory Model consists of a set of processors (with local memory)  connected by
links under some topology, and communication is carried out trough  message passing.

In attempting to address the issues related to the communication cost in distributed
memory systems, a couple of models merit discussion: the {\em delay model}, in which the delay on the communication between any two processors, no matters their distance in the network~\cite{Papadimitriou} is captured. This model has been widely used to represent distributed memory systems, incorporating issues like the heterogeneity of processors ~\cite{Sena}. 

The absence of a standard model of parallel computation influenced many researchers to work on the attempt to establish a bridge between parallel applications and parallel machines. Valiant \cite{Valiant}  defined the Bulk-Synchronous Parallel (BSP) model, which represents a set of processing elements, their speed, the time between two synchronization events, which characterizes  a  superstep. It is during each superstep that computation of tasks and message delivery between processors are supposed to be carried out. In a continuous search for more accurate models and with the advent of computer clusters, studies led to the specification of  HBSP~\cite{Williams} to model the heterogeneity of the processors, concerning their speeds and capacities.

Due to the emergence of network of workstations as high performance environment,  the LogP model  \cite{Culler} was  proposed to be a computational model in which global characteristics of parallel architectures are represented, such as number of processing elements, latency on the transmissions, gap between subsequent messages and overhead on the sending and receiving of messages. The key issues stated in the model were related to communication and non-synchronous computations. Following this work, other extended LogP models were proposed, as for example, in the LogGP Model \cite{Alexandrov},  the gap associated with the sending of long messages was  represented more accurately, while in the LogGPS \cite{Ino}, the cost associated with the necessary synchronization  when sending a long message under the MPI library is also captured. LoPC \cite{Frank} addresses contention problem that arises when sending messages in multiprocessors, i.e\ considers the sharing of global memory between processors. Regarding the point-to-point communication (i.e. send messages), which requires moving data from the source process local memory to the target process local memory, the models $Log_{n}P$ and $Log_{3}P$ are proposed in \cite{Cameron}. The model includes middleware costs into the representation of distributed communication.

Note that, on the comparison between the BSP and LogP models researchers have classified BSP as a suitable abstraction for parallel application development, while LogP offers a better resource management \cite{Bilardi,Ramachandran}. 

Following the advent of computer cluster, \cite{Tam,Mendes,Cameron} captured more precisely the sending and receiving overheads and latency. In their work, these costs depend on the size of the transmitted message, such that the costs being not the same for any transmission.

Yet, the architectural evolution has shown the benefits of a hybrid memory parallel system, where distributed memory computer are composed of machines with shared memory. Due to the actual technological advances, increasing execution performance of parallel applications on multicore systems become a reality. Still, further improvements are possible by properly characterizing such environments.

\subsection{ Multicore architectures - Models for distributed and shared memory architecture}

The actual trends for a cluster of multiprocessors are the multicore machines, which are connected by a network of some specific topology (as in a distributed memory multicomputer) thus defining a hybrid memory architecture that supports a hierarchical memory system. At the first level of the hierarchy, fine-grained applications could be performed reasonably well, while the second level supports efficiently coarse-grained applications. This ideal hierarchical parallelism modeling may be very powerful for the exploitation of the natural parallelism found in a great variety of applications.

Subsets of cores in a multicore machine may share different layers of memory levels. For example, usually, a small subset of cores shares L2 caches, while another subset of higher cardinality may share L3 caches, being the global memory shared by all the cores of the machine \cite{Badia,Chai,Savage2,Tu}.  The modeling of such memory hierarchy sharing  is still a challenge \cite{Savage}.

Multicores cannot be treated merely as shared memory processors like conventional symmetric multiprocessors (SMPs), mainly due to the design of multi-level cache hierarchies, which lead to a reduction on the memory bottleneck. Therefore, application performance will potentially benefit with a proper modeling of this architecture, mainly parallel ones (either that share or exchange data via message passing).

Typically, in shared memory models, the sharing happens for all processors at the main memory level. However, multicore processors have a varying degree of caches sharing at different levels. The \textit{Unifield Multicore Model} (UMM) proposed in \cite{Savage2} assumes that sets of cores share first-level caches, which in turn share second-level caches and that the cache capacity is the same for all caches at a given level. Also, in this work, lower bounds are derived for numerical application, but distributed memory is not account. 

Memory hierarchy should be captured among three levels of communication in a multi-core cluster: intra-processor, when communication is held between two cores on the same processor; inter- processor, when communication is carried out across processors but within the same machine; and inter-machine, between two cores on different machines. For the same message size, \cite{Ortega,Song2} captured distinct communication costs when communication is held between different levels. More specifically, \cite{Song2} defines an analytical model that considers different memory levels, and specifies an affinity degree between threads, depending on the data amount exchanged between them. Threads with higher affinity should be allocated to cores that shares lower memory level (i.e.\ cache), in order to avoid higher communication costs when these threads are in distinct processors. In this case, recall that main memory is being shared. Nonetheless, this model does not consider memory size, and at the end, too many threads can be allocated to share the same cache, and as a consequence the amount of cache miss might be increase \cite{Chai,prasanna_grupos}. The importance of accurately representing the communication costs depending on the memory hierarchy regarding the evaluation carried out by~\cite{Chai} on various applications, suggested that intra and inter-processor communication is as important as inter-machine communication, and data locality techniques that avoid memory contention must be designed to improve application performance.

%------------------------------------------------------------------------- 

%------------------------------------------------------------------------- 
%%%% the experiments with the synthetic application
%------------------------------------------------------------------------- 

\subsection{The application model}

The {\em application model}  defines the relevant characteristics related to the application performance, which is usually represented by directed acyclic graphs (DAGs), denoted by $G=(V,E,\varepsilon,\omega)$, where: the set of $n$ vertices $V$ represents {\em tasks}; $E$, the precedence relation among them; $\varepsilon(v)$ is the amount of work or computational weight associated with task $v \in V$; and $\omega(u, v)$ is the amount of transmitted data or communication weight associated with the edge $(u,v) \in E$, representing the amount of data units transmitted from task $u$ to $v$.  Also, since in the target system being considered in this work, memory sharing is closely related to the application performance, the amount of data required by task $v$ must be depicted and is represented by $\mu(v)$. 

%------------------------------------------------------------------------- 
\section{On Modeling Multicore Clusters}

In order to identify the influence of the relevant architectural characteristics on the application performance on multicore systems, a simple application, based on \cite{servet,Saavedra} was applied. This application consists of two tasks that execute two well defined phases: computation and communication. The computation phase corresponds to a two nested loops that scans a vector of integers in steps of 1K bytes, so that hardware prefetching is avoided, since the step size is bigger than any cache line and also the cache size is a multiple of this step size ~\cite{Saavedra}. The manner in which the vector is accessed also avoids further optimizations carried out by the compiler, as discussed in ~\cite{servet}.

The communication phase consists of the sending of a message from one task to another, such that one task executed a sending command, while the other a receiving. The way that this communication is actually carried out depends on whether the communicating tasks are allocated: if they are on the same machine, communication is held via shared memory, where semaphores are used to prevent race condition. Otherwise, a message is effectively transmitted.

All the experiments described in this section were executed in at least two machines of the multicore cluster RIO with  Gigabit interconnection network. Each machine is a quad-core Intel Xeon E5410 - Harpertown, each core with a private L1 cache of 64KB, and every two cores share a L2 12MB cache in each one of the two processors of a machine. All the four cores have a uniform access to a 16MB main memory module. Cent OS 5.3 is the operating system with kernel version 2.6.18. The application is implemented with Intel MPI version 4.0.0.028 and Posix was used to create threads. The PAPI tool \cite{papi} was used to collect and evaluate the execution performance of the application.

In order to evaluate the influence of memory sharing during the execution of the  application tasks on the machine cores, the following allocation was set:

\renewcommand{\theenumi}{\roman{enumi}{.}}
\renewcommand{\labelenumi}{\roman{enumi}{.}}

\begin{enumerate}
\item two tasks were allocated to the {\em same core}, and consequently, accessing the same cache (SC);
\item two tasks allocated to {\em different cores}, but sharing the same cache (SCM);
\item two tasks allocated to cores that do not share the same cache, but share the main memory (SMM);
\item two tasks allocated to cores of distinct machines (DM), where the global memory of each machine is not shared;

\end{enumerate}

\renewcommand{\theenumi}{\arabic{enumi}{.}}
\renewcommand{\labelenumi}{\arabic{enumi}{.}}

Let $\mu(v)$ be the vector size allocated by a task $v$  during the computation phase, as described above.  In order to enforce a given allocation of a task to a specific core, the system call {\sf set\_affinity()} \cite{Sadaf,Dongarra} was used and also, application tasks and system processes were not executed on the same core.

\subsection{Computation Phase Tasks}

In this experiment, two independent tasks $v_1$ and $v_2$, which do not communicate, were allocated under the SC, SCM and SMM allocation only. Note that in this experiment,  each task only performs the two nested loops that scans the vector and no sending and receiving was specified.   

\begin{figure*}[!ht]
\centering 
 \begin{center}
 \includegraphics[width =\linewidth, scale=0.8]{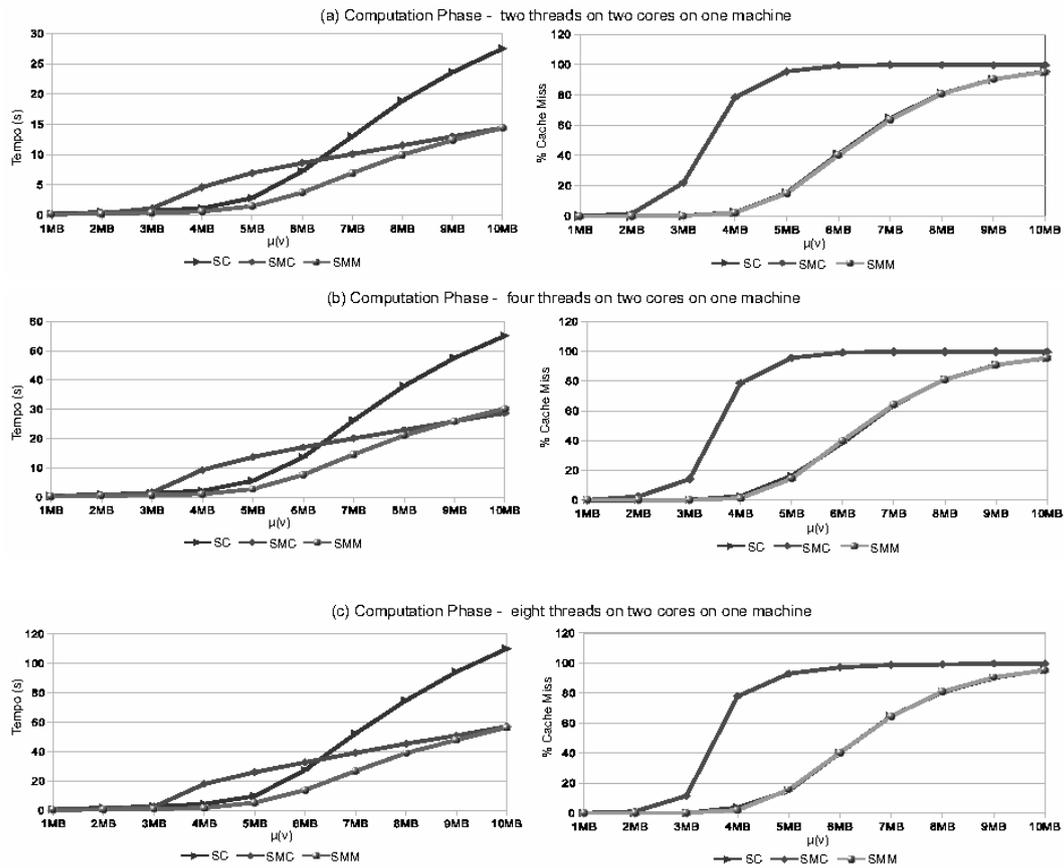}  
 \caption{Analysis on the execution of (a) 2, (b) 4 and (c) 8 threads in two cores on one machine.}
 \label{faseComputacao}
 \end{center}
\end{figure*}

It was observed that, even though the amount of data of both tasks is less than the cache capacity, the allocation SC was the one that produced the worst execution times, as shown in Figure \ref{faseComputacao}. This is due to the fact that, in the case of SC, both tasks were competing for the same computational resource. In the case that the amount of data allocated by each task is between 3MB and 6MB, the allocation SMM provided the best performance, since \ even when the whole amount of date for both tasks $\mu (v_1)$ and $\mu(v_2)$ was more than the cache capacity, the number of cache misses degraded the execution performance in the case of SCM. Therefore, it is better to use SMM, but on the same machine, since L2 cache is not shared. In the SMM allocation, the time can be reduced in $14.88$\%, when comparing with the SCM allocation (distinct cores, but same cache).  As a consequence for $\mu (v_i) > 6MB$ both tasks need more than the cache capacity and obviously, the number of global memory accesses highly increases.  

It is important to note that, although the execution time for two tasks executed on the same core (SC) is worse than the other two allocations (SCM and SMM), the relative number of cache misses are smaller than those for SCM and SMM, as seen in Figure \ref{faseComputacao} (a). This is in fact due to the sharing of computational resource rather than the cache memory. 

Experiments with four and eight threads, also on two cores of the same machine,  were also performed, whose results can be seen in Figure \ref{faseComputacao} (b) and (c). Evaluating the curves, one can see that although the overall execution time increased since more threads were allocated to the same core, the same behavior as the previous experiment was detected, where SMM leaded to the best performance, mainly for $\mu (v) \geq 3MB$, while, SC was always worse. Note that, the number of cache misses followed the same pattern as the one observed in Figure \ref{faseComputacao} (a).

\begin{figure*}[!ht]
 \begin{center}
 \includegraphics[scale=0.6]{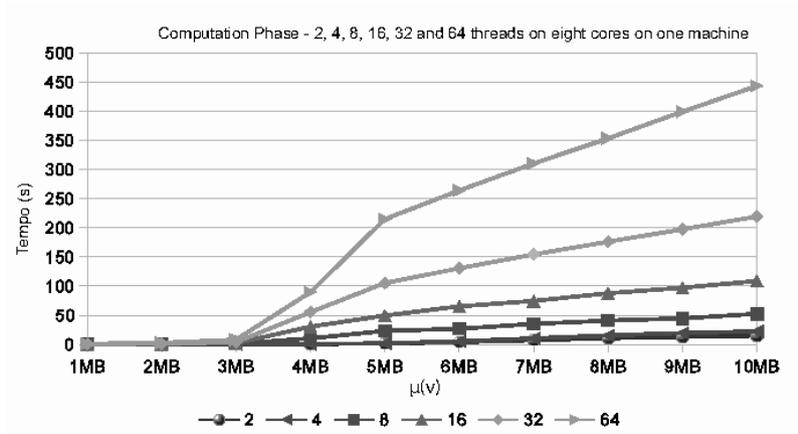}  
 \caption{Analysis on the execution from 2 to 64 threads in eight cores on one machine.}
 \label{faseComputacaoRam}
 \end{center}
\end{figure*}

The results of another experiment can be seen in Figure \ref{faseComputacaoRam}, where the number of threads $n = 2, 4, 8, 16, 32, 64$ was executed on one machine, being divided between its cores. In the case of $n < 8$, no more than one thread was executed per core, avoiding therefore, the SC allocation. For $n = 2, 4$, no cache sharing was held. 

Some interesting conclusions can be withdrawn from this experiment. For more than 3MB per thread, the higher is the number of threads, the higher is the application execution time, suggesting that  it is not worth executing more than one thread per processor. The bottom line is to allocate a number of threads per machine that does not fill the cache capacity.

\subsection{Communication phase}
\label{modelcomphase}

\begin{figure*}[!ht]
 \begin{center}
 \includegraphics[scale=1.0]{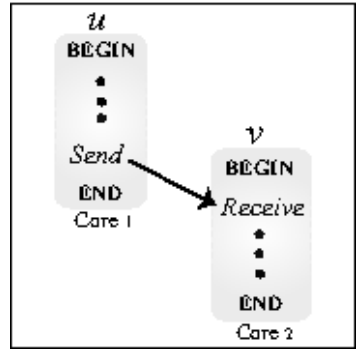}  
 \caption{Computation phase - using more than two cores}
 \label{exemplocomunicacao}
 \end{center}
\end{figure*}

In this experiment, the application consists of one computation and one communication phases, as seen in Figure \ref{exemplocomunicacao}. It consists of two tasks or threads, $v$ and $u$, allocated under the SC, SCM, SMM and DM (to evaluate the communication influence also between distinct machines) allocation, respectively.   It is important to note that whatever the allocation considered, the threads are practically not being executed in parallel due to the application topology.  As shown in Table \ref{send_time}, the communication phase time with threads allocated to the same machine is practically negligible. 

The experiment was repeated by executing  ten threads in two core under the SC, SCM, SMM and DM allocations. The application starts with one thread executing on one core its computation phase, and then sends a message to the another thread allocated to another core. This thread, after receiving the message and executing its computation phase, sends a message to another thread also allocated to the first core. This patterns follows for remaining threads, which upon receiving a message, execute the computation phase and then send a message to a different thread. Remark that a thread terminates as soon as it sends a message.

\begin{figure}[!h]
 \begin{center}
 \includegraphics[scale=0.6]{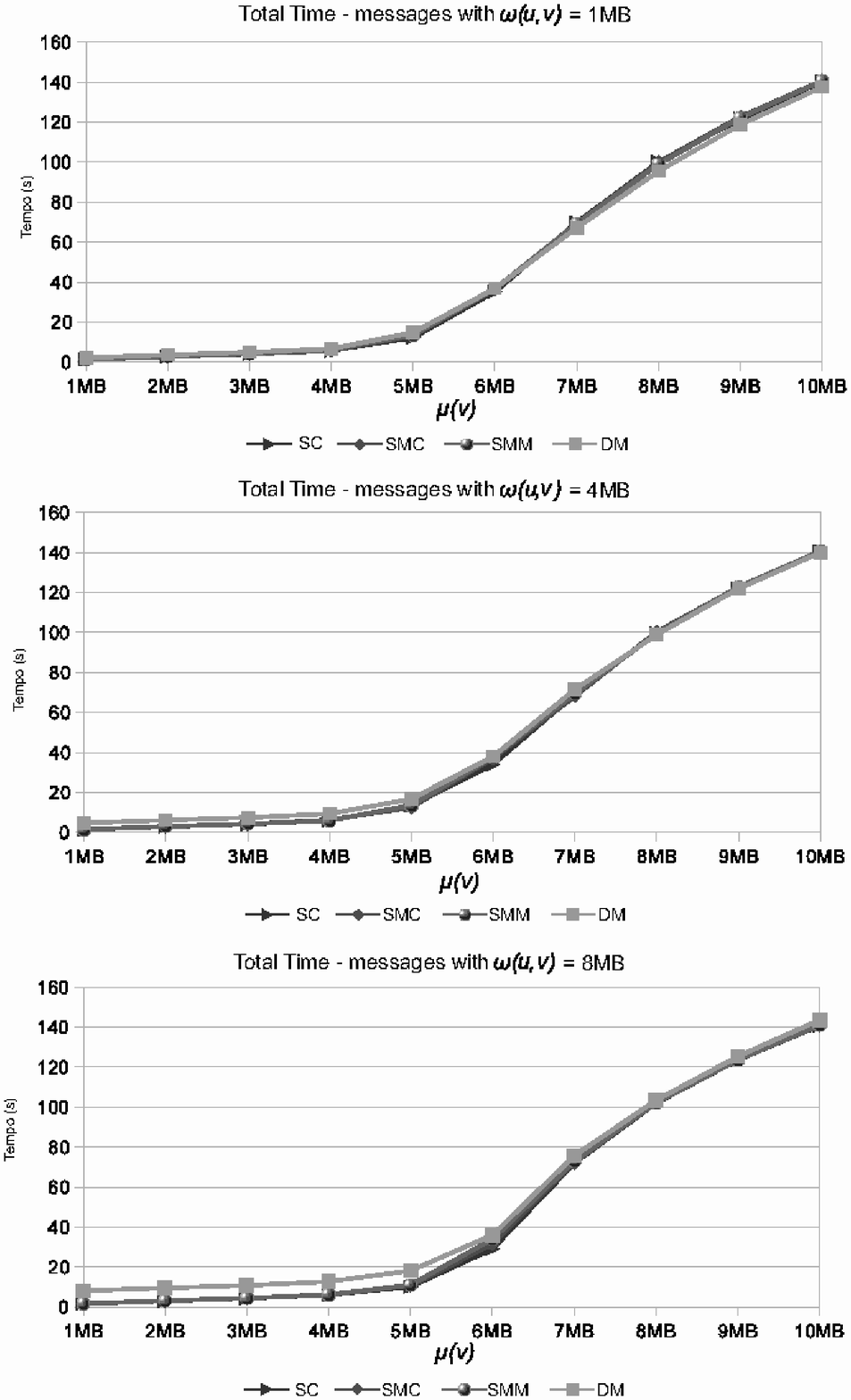}  
 \caption{ Total execution time (s) of ten threads under the SC, SCM, SMM and DM allocation}
 \label{faseComunicacao}
 \end{center}
\end{figure}

The results of this last experiment are shown in Table \ref{total_time_ten} and in Figure \ref{faseComunicacao}, and they represent the total execution times, with a varying message size $\omega(u,v) = 1MB, 4MB$ and $8MB$, respectively, where the $x$-axis of each graph corresponds to the vector size $\mu(v)$ of task $v$. From these results, one can note that when the vector size $\mu(v)$ is less then 6MB, the worst results are those produced by the DM allocation, since the communication cost associated with the message transmission inside a same machine is the smallest one. However, for $\mu (v) \geq 6MB$, the contention memory problem may arise, depending on the size of the message being sent. The overall execution time is slightly better for DM  when the messages are smaller than 8MB, that is $\omega(u,v) < 8MB$. Remark that a 8MB message cannot be considered a very long one considering the nowadays network performance. 

\begin{table*}[h!]
\tiny
\centering
\caption{Sending Time}
\begin{tabular}{c|r|r|r|r|r|r|r|r|r|r}
\hline
\backslashbox{Alloc}{$\mu(v)$} & \multicolumn{1}{c|}{\textbf{1MB}} & \multicolumn{1}{c|}{\textbf{2MB}} & \multicolumn{1}{c|}{\textbf{3MB}} & \multicolumn{1}{c|}{\textbf{4MB}} & \multicolumn{1}{c|}{\textbf{5MB}} & \multicolumn{1}{c|}{\textbf{6MB}} & \multicolumn{1}{c|}{\textbf{7MB}} & \multicolumn{1}{c|}{\textbf{8MB}} & \multicolumn{1}{c|}{\textbf{9MB}} & \multicolumn{1}{c|}{\textbf{10MB}} \\ \hline
\multicolumn{ 11}{c}{\textbf{$\omega(u, v)$ = 1MB message}} \\ \hline
SC & 0.000002 & 0.000001 & 0.000002 & 0.000002 & 0.000001 & 0.000001 & 0.000001 & 0.000001 & 0.000001 & 0.000001 \\ 
SCM & 0.000001 & 0.000001 & 0.000001 & 0.000001 & 0.000001 & 0.000001 & 0.000001 & 0.000001 & 0.000001 & 0.000001 \\ 
SMM & 0.000001 & 0.000001 & 0.000001 & 0.000001 & 0.000001 & 0.000001 & 0.000001 & 0.000001 & 0.000001 & 0.000001 \\ 
DM & 0.074050 & 0.074154 & 0.074178 & 0.074095 & 0.074185 & 0.074185 & 0.074200 & 0.074115 & 0.074162 & 0.074194 \\ \cline{ 1- 11}
\multicolumn{ 11}{c}{\textbf{$\omega(u, v)$ = 4MB message}} \\ \hline
SC & 0.000001 & 0.000002 & 0.000001 & 0.000002 & 0.000001 & 0.000001 & 0.000001 & 0.000001 & 0.000002 & 0.000001 \\ 
SCM & 0.000001 & 0.000001 & 0.000001 & 0.000001 & 0.000001 & 0.000001 & 0.000001 & 0.000001 & 0.000001 & 0.000001 \\ 
SMM & 0.000001 & 0.000001 & 0.000001 & 0.000001 & 0.000001 & 0.000001 & 0.000001 & 0.000002 & 0.000001 & 0.000001 \\ 
DM & 0.345687 & 0.345788 & 0.345731 & 0.345706 & 0.345729 & 0.345763 & 0.345771 & 0.345788 & 0.345780 & 0.345723 \\ \cline{ 1- 11}
\multicolumn{ 11}{c}{\textbf{$\omega(u, v)$ = 8MB message}} \\ \hline
SC & 0.000001 & 0.000001 & 0.000001 & 0.000002 & 0.000001 & 0.000002 & 0.000001 & 0.000002 & 0.000001 & 0.000001 \\ 
SMC & \multicolumn{1}{c|}{0.0000008} & 0.000001 & 0.000001 & 0.000001 & 0.000001 & 0.000001 & 0.000001 & 0.000001 & 0.000001 & 0.000002 \\
SMM & \multicolumn{1}{c|}{0.0000014} & 0.000001 & 0.000001 & 0.000001 & 0.000001 & 0.000001 & 0.000001 & 0.000001 & 0.000001 & 0.000002 \\ 
DM & \multicolumn{1}{c|}{0.7022743} & 0.702273 & 0.702338 & 0.702368 & 0.702299 & 0.702308 & 0.702349 & 0.702374 & 0.702367 & 0.702362 \\ \hline

\end{tabular}
\label{send_time}
\end{table*}

\begin{table}[htbp]
\tiny
\centering
\caption{Total Time  - ten threads in one machine}
\begin{tabular}{c|r|r|r|r|r|r|r|r|r|r}
\hline
\backslashbox{Alloc}{$\mu(v)$} & \multicolumn{1}{c|}{\textbf{1MB}} & \multicolumn{1}{c|}{\textbf{2MB}} & \multicolumn{1}{c|}{\textbf{3MB}} & \multicolumn{1}{c|}{\textbf{4MB}} & \multicolumn{1}{c|}{\textbf{5MB}} & \multicolumn{1}{c|}{\textbf{6MB}} & \multicolumn{1}{c|}{\textbf{7MB}} & \multicolumn{1}{c|}{\textbf{8MB}} & \multicolumn{1}{c|}{\textbf{9MB}} & \multicolumn{1}{|c}{\textbf{10MB}} \\ \hline
\multicolumn{ 11}{c}{\textbf{$\omega(u, v)$ = 1MB message}} \\ \hline
SC & 1.3374 & 2.7777 & 4.0447 & 5.7713 & 12.2419 & 35.2310 & 69.7924 & 100.1719 & 121.5122 & 140.4271 \\ 
SCM & 1.3414 & 2.7778 & 4.0471 & 5.7029 & 13.3135 & 36.4155 & 69.2166 & 99.9814 & 122.7039 & 140.8029 \\ 
SMM & 1.3538 & 2.8219 & 4.1223 & 6.4391 & 13.5027 & 36.7180 & 69.2976 & 98.9662 & 122.3274 & 140.6743 \\ 
DM & 2.1571 & 3.5942 & 4.9074 & 6.6670 & 14.7997 & 36.8324 & 67.1933 & 95.5452 & 118.8403 & 137.5865 \\ \hline

\multicolumn{ 11}{c}{\textbf{$\omega(u, v)$ = 4MB message}} \\ \hline

SC & 1.4811 & 2.9270 & 4.1948 & 5.9055 & 13.6012 & 34.0697 & 69.1819 & 99.6317 & 122.0155 & 140.5231 \\ 
SCM & 1.4850 & 2.9283 & 4.1952 & 6.2532 & 12.5493 & 37.0250 & 68.2627 & 100.3262 & 122.6561 & 140.5614 \\ 
SMM & 1.5016 & 2.9637 & 4.2477 & 5.9298 & 13.5898 & 37.7790 & 69.5724 & 99.4298 & 122.9522 & 140.6809 \\ 
DM & 4.7281 & 6.1448 & 7.4163 & 9.3374 & 16.7040 & 38.0295 & 71.5360 & 98.9790 & 121.9815 & 140.0319 \\ \hline

\multicolumn{ 11}{c}{\textbf{$\omega(u, v)$ = 8MB message}} \\ \hline
SC & 1.6758 & 3.1227 & 4.3917 & 6.0852 & 10.1448 & 29.0173 & 71.8652 & 101.9864 & 123.6068 & 140.8053 \\ 
SCM & 1.6788 & 3.1228 & 4.3935 & 6.2708 & 10.7288 & 31.3465 & 72.1163 & 102.8356 & 124.0472 & 140.7693 \\ 
SMM & 1.6958 & 3.1611 & 4.4446 & 6.1554 & 11.0589 & 33.9579 & 72.8210 & 102.1969 & 123.6122 & 141.1606 \\ 
DM & 8.1402 & 9.5474 & 10.8190 & 12.7849 & 18.1668 & 36.0094 & 75.6678 & 103.3968 & 125.3568 & 143.6539 \\ \hline
\end{tabular}
\label{total_time_ten}
\end{table}

%\begin{figure*}[!hb]
% \begin{center}
% \includegraphics[scale=0.5]{comunicacao10t.eps}  
% \includegraphics[scale=0.5]{comunicacao10T_4mb.eps} 
% \includegraphics[scale=0.5]{comunicacao10t_8mb.eps}
% \caption{ Total execution time (s)}
% \label{faseComunicacao}
% \end{center}
%\end{figure*}

%\begin{figure}[!h]
% \begin{center}
% \includegraphics[scale=1]{comunicacao10T_4mb.eps}  
% \caption{ Total execution time (s)}
% \label{faseComunicacao}
% \end{center}
%\end{figure}
%
%\begin{figure}[!h]
% \begin{center}
% \includegraphics[scale=1]{comunicacao10t_8mb.eps}  
% \caption{ Total execution time (s)}
% \label{faseComunicacao}
% \end{center}
%\end{figure}

%------------------------------------------------------------------------- 
%%%% the Model MCM

\subsection{Multicore Clusters Model - MCM}
\label{modelodaarquitetura}

In the light of the above analysis, this section describes the proposed Multicore Cluster Model (MCM), where a multicore cluster $CM = \{ M_{0},M_{1},M_{2},\ldots ,M_{m} \}$  is set of $m$ machines, where each machine $M_{i}$, $1 \le i \le m$ consists of a set of $p$  processors  $P_{i} = \{P_{(i,0)},P_{(i,1)},P_{(i,2)},\ldots ,P_{(i,p)}\}$. In turn, each processor $P_{(i,j)}$ consists of a set of $c$ cores, being each one denoted by $C_{(i,j,k)}$. 

Cores in the machine $M_i$ share the global main memory, $gm_i$, with capacity $gmc_{i}$ and cores in the processor $P_{(i,j)}$ share a cache memory in a given level. Each processor $P_{(i,j)}$ in each machine $M_i$ has a set of $l$ cache memories  $CM_i= \{cm_{(i,j,0)}, cm_{(i,j,1)} ,\ldots , cm_{(i,j,l)} \}$. The capacities of each cache $cm_{(i,j,k)}$ is denoted by $cmc_{(i,j,k)}$, such that $cmc_{(i,j,k)} < gmc_{i}$, i.e., \ the capacity of the cores are smaller than the global memory one. 

Every two cores $C_{(i,j,k1)}$ and  $C_{(i,j,k2)}$, which share the cache memory $cm_{(i,j)}$ are called neighbor cores. Also, all the cores in a machine share the global memory $gm_i$. 

All the cores in the machine $M_i$ have the same {\em computational slowdown index} $csi_i$, which is an estimation of the computational power of each core in $M_i$, as defined in \cite{Sena}. Therefore, MCM models homogeneous cores inside a machine, but the machines are not necessarily homogeneous. Thereby, the sole execution time associated with task $v$ in a core, say, $C_{(i,j,k)}$, is then $et (v, C_{(i,j,k)}) = csi_i \times \epsilon_{i_{j_k}}$.

Concerning the cache influence on the application performance,  this work defines the worst case execution time of a task $v$ on a given core $C_(i,j,k)$ due to the number of cache misses that might occur, which depends on the amount of memory already allocated. Hence, the execution time of task $v$ is established not only by the computational slowdown index, but also, the amount of data already allocated to $cm_{(i,j,k)}$ and the main memory $gm_i$.

An edge $(u,v)$ represents the dependency between tasks $u$ and $v$ and also, the exchange of information between them, whose amount is given by $\omega (u,v)$. The communication time to transmit this data between two machines, say, $M_i$ and $M_j$ is then $ct ((u,v), M_i, M_j) = \omega (u,v) \times lat (M_i, M_j)$, where $lat (M_i, M_j)$ is the communication latency associated with the link between $M_i$ and $M_j$. 

Considering the previous tests related to the communication phase, it is considered, in MCM, that the communication cost inside a machine is negligible.

\subsection{A Load Balance Model}
\label{schedModelSec}

Regardless of the computational time associated with the sole execution of an application task in a core, this time is actually influenced by the amount  of tasks that are being executed on neighbor cores. Let $\epsilon(v)$ be the computational weight of a task $v$, $\mu_{v}$ memory amount allocated when executing $v$, and $\omega(u,v)$ the communication weight from each one of the immediate predecessors $u \in pred (v)$. Suppose that $u$ is  allocated to core $C_{(i0, j0, k0)}$. The task $v$ is allocated  to $C_{(i1, j1, k1)}$, which is related to  $C_{(i0, j0, k0)}$ depending on the following conditions:

\begin{enumerate}

\item if $ (\mu_{u} + \mu_v) < cmc_{(i1, j1, k1)}$, the execution time of $v$ is the smallest one if either $i0 = i1$, that is, if the amount of data required by both $u$ and $v$ is smaller than the cache memory capacity, the computational time  will be the smallest if both tasks are allocated on the same machine but distinct cores, no matters if cache memory is shared or not. In the case $(u,v) \in E$, for $\omega(u,v) < LB_{msg}$, the total execution time of $v$ will be smaller if both tasks are executed in the same machine. \label{cond1}

\item if  $ (\mu_{u} + \mu_v) \geq cmc_{(i1, j1, k1)}$, the computation time of $v$ is smaller if  $i0 = i1$, $j0 \ne j1$ and $k0 \ne k1$ , that is, if the amount of data required by both $u$ and $v$ is more the cache capacity, the computation time of $v$ will be smaller if both tasks are executed on distinct cores of the same machine, but cache is not shared (non-neighbor cores). In the case $(u,v) \in E$, the amount of data to be transmitted should be considered:

\begin{enumerate}	
	\item if it is bigger than the cache size, the computation time of $v$ is smaller if  $i0 = i1$, that is, both tasks are executed in  the same machine. \label{cond2a}

	\item otherwise, the communication message is smaller than the whole cache, $v$ should be allocated to  a different machine, that is, $i0 \ne i1$. \label{cond2b}	
\end{enumerate}

Although conflicting, condition \ref{cond2a} and \ref{cond2b} relies on the fact that it is cheaper to send small messages via network than to keep locally. On the other hand, for long messages (in this work, no more than 8MB), memory contention for such messages is not as expansive as the communication time via network.   

\end{enumerate}

%------------------------------------------------------------------------- 

%------------------------------------------------------------------------- 

\section{Load Balance of a Parallel Branch-and-Bound based on MCM}
\label{}

In order to analyze and validate MCM, a load balance procedure based on the MCM model  was  developed  in the context of a parallel branch-and-bound ($PB\&B$) algorithm applied to the Set Partitioning  Problem.

Branch-and-bound is a widely used technique for solving NP-hard optimization problems. Such algorithms search the space of solutions following a tree enumeration. As the computations along the subtrees can be accomplished almost independently, they are considered to be well suited for parallelism.

There exists a variety of papers in the literature  that propose parallel branch-and-bound algorithms  or frameworks  to ease its development  for distributed \cite{Bauer, Bob, Cun2011, Lucia} and shared memory \cite{SYMPHONY, Bob, Estrada, Kim, PICO,PUBB, Rashid} architectures. However, to the best of our knowledge, few of them  explores both shared and distributed memory. Moreover,  they  do not consider the memory hierarchy of  multicore processors in their solutions \cite{Werneck}.

For a better understanding of this method, an introduction of the sequential $B\&B$ applied to the Set Partition Problem follows.

%-----------------------------------------------------

\subsection{Sequential $B\&B$ applied to the Set Partitioning Problem}

Given $n$ variables $x_1,\ldots,x_n$ with corresponding costs $c_1,\ldots,c_n$ and 0-1 coefficients
$a_{1j},\ldots, a_{nj}$, for $j = 1, \ldots, m$,
the Set Partitioning Problem (SPP) is the problem of assigning 0-1 values to these variables
such that $\sum_{i=1}^n a_{ij} x_i = 1$, for $j = 1, \ldots, m$, minimizing $\sum_{i=1}^n c_i x_i$.
Besides the many applications of this problem, the SPP is a problem of great interest because it is
a natural special case of integer programming.

\subsubsection{Lower Bound}

A straightforward lower bound on the optimal solution for this problem can be calculated by solving of
its continuous relaxation

\begin{alignat}{3}
\mbox{Minimize} \quad & \sum_{i=1}^n c_i x_i & \label{eq:primal1} \\
\mbox{subject to} \quad & \sum_{i=1}^n a_{ij} x_i = 1 \quad & j=1,\ldots,m \label{eq:primal2} \\
& x_i \geq 0, & i=1,\ldots,n \label{eq:primal3}
\end{alignat}

\noindent
or its dual

\begin{alignat}{3}
\mbox{Maximize} \quad & \sum_{j=1}^m \pi_j & \label{eq:dual1} \\
\mbox{subject to} \quad & \sum_{j=1}^m a_{ij} \pi_j \leq c_i \quad & i=1,\ldots,n. \label{eq:dual2}
\end{alignat}

\noindent
where the $\pi_j$ variables may assume either positive or negative values.

The advantage of using (\ref{eq:dual1}-\ref{eq:dual2}) to instead of (\ref{eq:primal1}-\ref{eq:primal3}) is that optimality is not necessary. In our branch-and-bound procedure, we use the following heuristic to calculate a feasible dual solution that approaches its optimal solution in a reduced computational time.

Our dual heuristic repeats two main steps by a fixed number of iterations. The first step, that we call the {\it forward step}, consists of increasing the $\pi_j$ values as much as possible. Then, in the {\it backward step}, it reduces some $\pi_j$ values while increasing others aiming to be able to improve the lower bound in the next forward step. Hence, the backward step is not executed in the last iteration.

The forward step is also divided into a number of iterations. In each iteration, the same value $\Delta_1$ is added to each $\pi_j$ that does not belong to a saturated constraint, i.e., $\Delta_1$ is added to $\pi_j$ if and only if $\sum_{j=1}^m a_{ij} \pi_j < c_i$ for all $i$ such that $a_{ij} = 1$. Since $\Delta_1$ is chosen as the maximum value that will keep all constraints (\ref{eq:dual2}) satisfied, at least one new constraint becomes saturated upon every iteration. The forward step stops when no more $\pi_j$ variables can be increased. This step is part of a well-known approximation algorithm for the SPP \cite{Hochbaum}.

In the backward step, the value of $\pi_j$ is decreased by $\Delta_2 (\alpha_j - 1)$, for some $\Delta_2$, where $\alpha_j$ is the number saturated constraints where $\pi_j$ has a non-zero coefficient. If $\alpha_j = 0$, then $\pi_j$ is increased by $\Delta_2$. The value of $\Delta_2$ chosen so that the current lower bound is multiplied by a given factor $\theta$. We use $\theta = 0.5$ in the first iteration of the root node and $\theta = 0.3$ in the first iteration of the remaining nodes. After each iteration, $\theta$ is multiplied by $0.7$. We perform 10 iterations in the root node and 5 in the remaining nodes.

\subsubsection{Branching}

We do branching on the constraints (\ref{eq:primal2}).
For a selected row $j$, we create one branch for each $i$ with $a_{ij} = 1$ where the variable $x_i$ is fixed to one.

One important characteristic of the SPP is that each child node can be substantially smaller than its parent.
Whenever a variable $x_i$ is fixed to one, every variable $x_k$ such that both $a_{ij} = 1$ and $a_{kj} = 1$ for some $j$ can be fixed to zero. Then, every constraint (\ref{eq:primal2}) where $x_i$ has a non-zero coefficient can be removed. In our method, the remaining constraints inherit the values of $\pi_j$ from the parent node.

Next, we describe the criterion used to select a constraint $j$ for branching. Let $\delta_i$ be the number of constraints $\ell$ such that $a_{i\ell} = 0$. We select the constraint $j$ with the smallest value of $\sum_{i \in \{1,\ldots,n\} \atop a_{ij}=1} \delta_i$, which represents the total number of constraints in all child nodes that would be created.

In order to find feasible solutions earlier, we process the child nodes in a non-decreasing order of
$(c_i - \sum_{j=1}^n a_{ij} \pi_j) / \delta_i$.
The branch-and-bound tree is traversed in a depth-first search fashion.

A more sophisticated and effective dual heuristic for the set partitioning problem has been proposed recently in \cite{dual}. However, we decided to use our own heuristic because it is simpler and achieves comparable lower bounds for the instances used in our experiments.

%-----------------------------------------------------

\subsection{Parallel Branch-and-Bound applied to the Set Partitioning Problem - $PBB_{SPP}$}

The parallel algorithm was grounded on the perviously described Branch-and-Bound algorithm for the Set Partitioning Problem. The $PBB_{SPP}$ incorporates interesting characteristics in relation to memory management. At first, it does not generate a binary tree, and actually, the number of subtrees generated by each node  can vary a lot.  Also, nodes execution times are usually very small, on average between 0.001 to 0.006 seconds, depending on the instance.  However, many of these nodes can need a larger amount of  memory (this necessary amount is referred as node size). 

Table \ref{nodes} presents  information about node sizes in bytes  and their corresponding times in seconds. For four instances, it is shown the five smallest (five first lines of each instance) and the five largest node sizes (the five remaining lines of each instance) for four different instances.  The table also presents the associated levels  (distance from the root) of those nodes in the $B\&B$ tree. The instances used in the tests were randomly generated. The two first numbers of the instance name refer to the quantity of items and sets, respectively. The remaining information refers to the probability that items appear in the set, followed by the seed of randomness.

It can be observed that all executions times of the nodes are very small. It is important also to note that the lowest level nodes demand much more memory than the highest level ones. Since the quantity of saturated constraints are smaller in lowest level nodes than in the highest level ones.

\begin{table*}[htbp]
\centering
\footnotesize
\caption{ Example reporting the level, size and execution time of nodes for four instances of SPP}
\begin{tabular}{|r|r|r||r|r|r|}
\hline
\textbf{Level } & \textbf{Node Size (Bytes)} & \textbf{Node Execution Time} & \textbf{Level } & \textbf{Node Size (Bytes)} & \textbf{Node Execution Time} \\ \hline
\multicolumn{3}{|c||}{\textbf{I90-400-0.03}}  & \multicolumn{3}{c|}{\textbf{I100-500-0.03}} \\ \hline
17 & 448 & 0.001 & 25 & 420 & 0.001\\ 
17 & 452 & 0.001 & 22 & 424 & 0.105 \\ 
18 & 456 & 0.085 & 22 & 448 & 0.001 \\ 
18 & 480 & 0.001 & 26 & 472 & 0.001 \\ 
16 & 516 & 0.001 & 27 & 480 & 0.005 \\ 
2 & 9092 & 0.005 & 1 & 9836 & 0.096 \\ 
3 & 9208 & 0.005 & 2 & 10068 & 0.006 \\ 
1 & 9216 & 0.076 & 1 & 10452 & 0.001 \\ 
2 & 9344 & 0.005 & 2 & 10888 & 0.006 \\ 
1 & 9436 & 0.062 & 1 & 11780 & 0.083 \\ \hline
\multicolumn{3}{|c||}{\textbf{I110-750-0.03}} & \multicolumn{3}{c|}{\textbf{I200-650-0.02-100}} \\ \hline
24 & 508 & 0.001 & 25 & 2608 & 0.001\\ 
24 & 516 & 0.012 & 19 & 2640 & 0.001 \\ 
29 & 528 & 0.001 & 23 & 2724 & 0.000 \\ 
29 & 532 & 0.034 & 22 & 2728 & 0.001 \\ 
24 & 532 & 0.026 & 23 & 2828 & 0.002 \\ 
2 & 17072 & 0.009 & 1 & 16960 & 0.003 \\ 
1 & 17476 & 0.006 & 1 & 17080 & 0.009 \\ 
1 & 17628 & 0.011 & 2 & 17400 & 0.011 \\ 
1 & 18492 & 0.003 & 1 & 18616 & 0.003 \\ 
1 & 18608 & 0.019 & 1 & 19004 & 0.008 \\ \hline
\end{tabular}
\label{nodes}
\end{table*}

\begin{figure*}[!ht]
 \centering
 \includegraphics[scale=0.85]{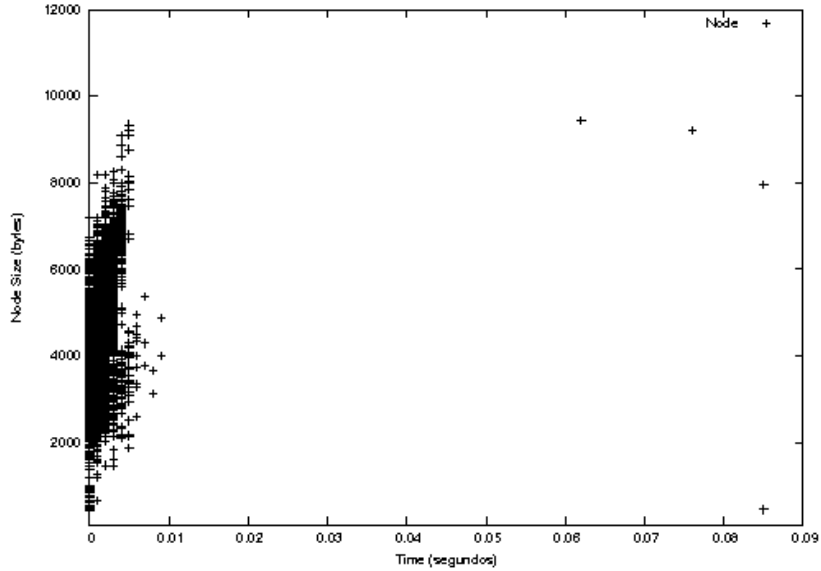}  
 \caption{Execution time and  nodes size for the I90-400-003 instance for the Set Partitioning Problem}
 \label{nodesgrafico}
\end{figure*}

Figure \ref{nodesgrafico} presents the execution time versus node size for the instance I90-400-0.03.  Most of the nodes spend very small  computation time, however their sizes vary a lot, from 50 Bytes to 9.6 KBytes.  All other instances analyzed in this work presented similar characteristics.

%-----------------------------------------------------

\subsubsection{The load Balance Framework}

The $PBB_{SPP}$ algorithm  assumes a static assignment of processes to machines such that exactly one process is assigned
to each physical machine $M_i$ of the cluster.  A process is composed of as many  threads as the number of cores of machine $M_i$, including a manager thread, $MT_i$, which  is responsible for generation of the remaining threads in  $M_i$, called workers,  and for communication with other machines of the cluster. At each core $C_{(i,j,k)}$, a worker thread denoted as $T_{(i,j,k)}$ executes the $B\&B$  tree nodes  until it  becomes idle, when then it initiates a procedure to obtain  new subtrees from other overloaded worker threads.
A unique leader thread (one leader per application), created on the machine $M_0$ and denoted as $LT$, is  responsible for starting and terminating the application.    

When a worker thread $T_{(i,j,k)}$ receives a node, it executes a branch-and-bound procedure which  generates other nodes that are kept in a local list of nodes $TL_{(i,j,k)}$. In accordance with a $B\&B$ parameter, each subtree can be  traversed either in breadth or depth way, which in turn can  affect  the size of the list $TL_{(i,j,k)}$. In both traverse schemes, the proposed load balance strategy respects the  associated cache size in accordance with the Condition \ref{cond1} of the {\em Load Balance Model} stated in Section \ref{schedModelSec}.  

The manager thread  $MT_i$  is responsible for requesting load  from another machine in the system. Let $M_j$ be a machine with overloaded threads. $MT_j$ removes parts of nodes from the lists of all threads, and sends them to $MT_i$, that requested load. If $MT_i$ is not able to obtain more load and all the respective threads are idle, it reaches its local termination condition, and informes this  to the leader of the application $LT$. The $PBB_{SPP}$ terminates when $LT$ receives the local termination condition from all manager threads  in the system.

\begin{figure*}[!ht]
 \centering
 \begin{center}
 \includegraphics[scale=0.57]{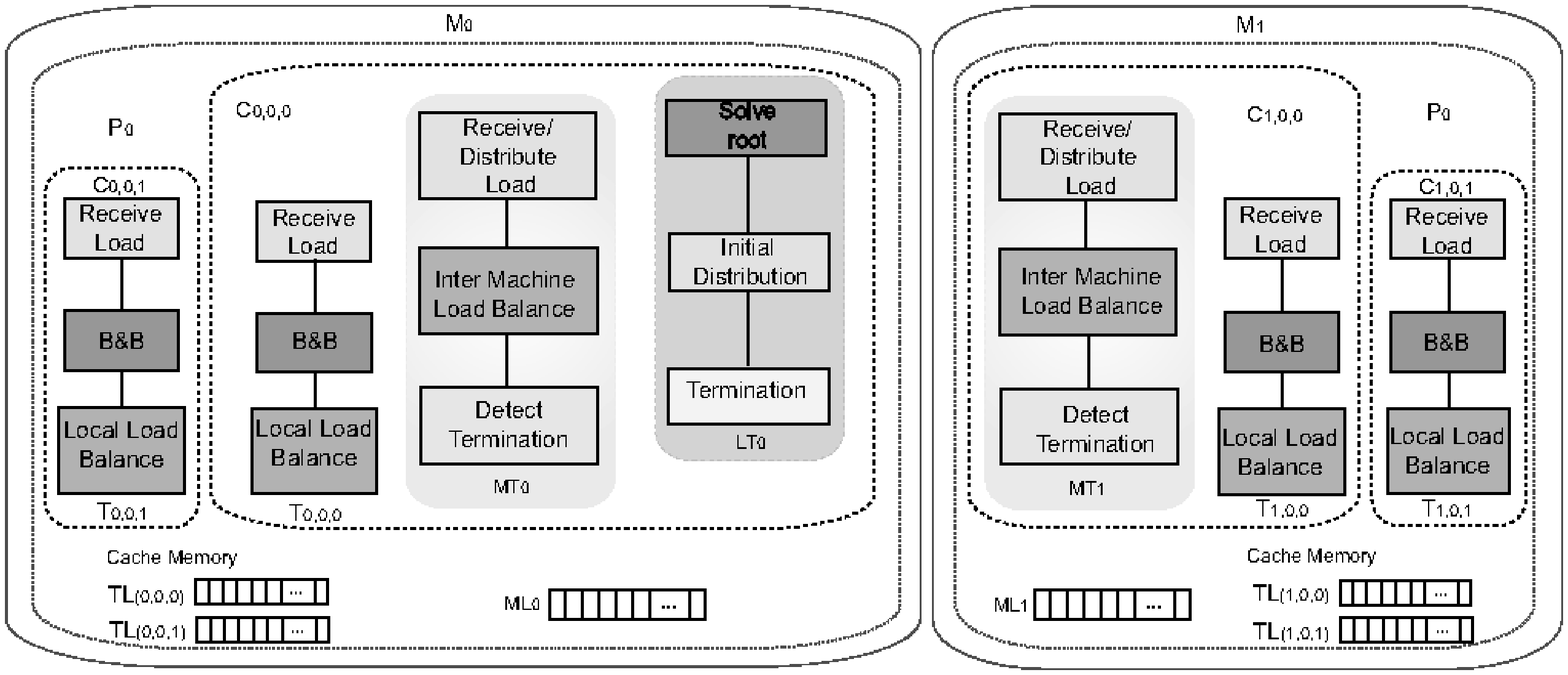}  
 \caption{ The Load Balance Framework on the target architecture}
 \label{load_balanceinter}
 \end{center}
\end{figure*}

Figure \ref{load_balanceinter} shows an example of two machines $M_0$ and $M_1$, each one with a processor, $P_{(0,0)}$ and $P_{(1,0)}$, respectively. Each processor has two cores $C_{(i,j,0)}$ and $C_{(i,j,1)}$ that share a common cache. The procedures executed by threads are represent by rectangles. Additionally, the figure shows the global lists, $ML_0$ and $ML_1$, used in the inter machine load balancing, and the local lists, $TL_{(0,0,0)}$, $TL_{(0,0,1)}$, $TL_{(1,0,0)}$ and $TL_{(1,0,1)}$ used in the load balancing among worker threads of the same machine.

\subsection{Load Balance Algorithms}
\label{bal}

The initial load distribution is performed by $LT$, which executes the root node of the parallel $B\&B$ tree.  As shown in  Algorithm \ref{distInicial}, the generated nodes are placed in the list $GL$ (line \ref{solve}). Considering that the function $numberNodes(GL)$  returns the quantity of nodes in this list,  $LT$ evenly shares the nodes among the  worker threads (line \ref{divide}), by sending $Load$ message (lines 2-10). 
\begin{algorithm}
\caption{Initial Distribution managed by $LT$}
\begin{algorithmic}[1]
\STATE {$GL$ = Solve(RootNode);} \label{solve}
\STATE {numNodes$ \gets \frac{numberNodes(GL)}{(m*p*c)}$;} \label{divide}
\FORALL {$i \gets 0$ to $m$} 
	\FORALL {$j \gets 0$ to $p$}
		\FORALL {$k \gets 0$ to $c$}
			\STATE {$Load \gets$ nodes($GL$, numNodes);} 
			\STATE {\textbf{Send} $Load$ \textbf{to} $T_{(i,j,k)}$;} 
 		\ENDFOR
	\ENDFOR
\ENDFOR
\end{algorithmic} 
\label{distInicial}
\end{algorithm}

In case of a thread $T_{(i,j,k)}$ does not receive any initial load (i.e. when $numberNodes(GL) < m*p*c$) or finishes executing its current load, it starts a load request procedure by executing  Algorithm \ref{ThreadSendLoadRequest}, which is actually performed whenever  $T_{(i,j,k)}$  becomes idle.

Upon finishing the execution of nodes of $TL_{(i,j,k)}$, the thread $T_{(i,j,k)}$ starts the load balance procedure in order to obtain nodes from other overloaded threads, whether exists, in the following order: a neighbor core at first (Algorithm \ref{ThreadSendLoadRequest});  secondly  from other threads of the  same machine since the neighbor threads are underloaded or even idle, i.e. their respective node lists are empty (Algorithm \ref{ThreadRecvNotLoad}, lines 6-9); and finally,  from  another machine, if the thread $T_{(i,j,k)}$ is not able to obtain load from other threads in  its own machine $M_i$ (Algorithm \ref{ThreadRecvNotLoad}, line 11).

Upon receiving a load request message,  $T_{(i,j,k)}$  sends a number of nodes from its local list as presented in line 2 of Algorithm \ref{ThreadRecvLoadRequest}  or send a message informing that its list is also empty as shown in line 6 of the same algorithm.

Concerning the manager thread, when it receives a number $NT$ of local load requests, it sends a request to another machine $M_x$ in a sequence of machines to be request, as presented in Algorithm \ref{MTRecvLoadRequestT}. When the other machine, $M_x$ answers the request by sending load, it shares the received load among the requesting load threads, as depicted in Algorithm \ref{MTRecvGLoadMT}. If, at last, it is not able of obtaining load from any other machine, it initiates the termination detection algorithm (in Algorithm \ref{MTRecvNotLoadMT}).

Finally, when a manager thread receives a load request from another machine, as portrayed depicted in Algorithm \ref{MTRecvLoadRequestMT}, it  tries to obtain load from all worker threads of its own machine. Upon receiving an answer from all worker threads, it forwards the total obtained load to the requesting machine by executing Algorithm \ref{MTSendGLoadMT}.

\subsection{Implementation issues of the Load Balance Framework}
\label{detalhes}

The proposed model MCM influenced the  implementation of the $PP_{BBSP}$ in many aspects, as described next.

In order to avoid a memory cache contention,  as  verified in the previous study, the list of nodes $TL_{(i,j,k)}$ managed  by each thread $T_{(i,j,k)}$ during the $PBB_{SPP}$  execution should not occupy more than its share, which is the total size of L2 cache memory divided by the number of threads that share it. Once this limit is reached,  a {\em recursive procedure} that performs depth-first-traversal  on the $B\&B$ tree is initiated.  The benefits of the depth-first-traversal  can  highly improve the  $B\&B$  performance.

In order to have useful  data at the last  level  cache when needed,  the local list of nodes  $TL_{(i,j,k)}$  for each thread was implemented, increasing the chances of processing them without accessing  the main memory. 
However, the so called false sharing might occur  when threads on different cores write to a shared cache line, but not at the same location. In this case, since the written locations are different, there is no real coherency problem, but the cache-coherency protocol sets the cache line to dirty, and when there exists an access request to the other location, the hardware logic will force a reload of a cache-line update from memory (even if not really necessary in logic terms). Frequent updates of the data in the shared-cache line could cause severe performance degradation.  In order to prevent this degradation each list of nodes was allocated in a different cache line.

Concerning the load balance procedure executed by $MT_i$, a global list  $ML_i$ of nodes  is also created at each machine  $M_i$. The Manager  Thread $MT_i$ disposes nodes  transferred from other machines, in  $ML_i$ and distributes this total load among the threads in $M_i$ that requested for load. The updating the global list, in both cases of storing and removing nodes in  $ML_i$  were implemented  with the  same rules of  the classical producer consumer  problem, guaranteeing that data were consistent and no deadlock occurred.

Considering that the load transferring  inside a machine, involves only two threads, a temporary list of nodes  is created with half of the nodes from the thread that contains load,  those nodes are removed  by the load requesting  thread. 

Remark that although the global list $ML_i$ can be larger than the available cache space, it will be used only when the internal load balance fails. As the next section shows, it happens very occasionally when compared with  the internal load transfers.  Note also that the time of communication among machines  is much higher than a node processing time and  transmitting very small loads can  increase the frequency of communication in the network. In  this case, the performance could be negatively affected.

\begin{center}

\fbox{
\def\baselinestretch{1}
\begin{minipage}{0.9\textwidth}
\centering
\begin{minipage}{0.8\textwidth}
\centering
\begin{algorithm}[H]
\caption{Load Request by $T_{(i,j,k)}$ when it becomes idle}
\begin{algorithmic}[1]
\IF {$TL_{(i,j,k)}$ = $\emptyset$}
\STATE{\textbf{Send} $LoadRequest$ \textbf{to} $T_{(i,j,l)}$;}
\ENDIF
\end{algorithmic}
\label{ThreadSendLoadRequest}
\end{algorithm}
\end{minipage}

\begin{minipage}{0.8\textwidth}
\begin{algorithm}[H]
\caption{When $T_{(i,j,k)}$ receives $NoLoad$ from $T_{(i,x,y)}$}
\def\baselinestretch{1}
\begin{algorithmic}[1] 
\IF{($x$ = $j$)} 
	\IF{($y+1 < c-1$)} 
		\STATE{$y++$;} 
    		\STATE{\textbf{Send} $LoadRequest$ \textbf{to} $T_{(i,j,y)}$;}  \COMMENT{Send request to another core in the same processor}
    \ELSE
     
    \IF{($j+1 < p-1$)}
    		\STATE{$j++$;} 
    		\STATE{$y \gets $0;}  
    		\STATE{\textbf{Send} $LoadRequest$ \textbf{to} $T_{(i,j,y)}$;} \COMMENT{Send request to another processor on the same machine} 
		\ELSE
		\STATE{\textbf{Send} $LoadRequest$ \textbf{to} $MT_i$;} \COMMENT{otherwise, forward request to manager thread}
    
    \ENDIF
    \ENDIF
\ENDIF
\end{algorithmic}
\label{ThreadRecvNotLoad}
\end{algorithm}
\end{minipage}

\begin{minipage}{0.8\textwidth}
\begin{algorithm}[H]
\caption{When $T_{(i,j,k)}$ receives $LoadRequest$ from $T_{(i,j,l)}$}
\def\baselinestretch{1}
\begin{algorithmic}[1]

  \IF {($TL_{(i,j,k)} \ne \emptyset$)}  
  	\STATE { numNodes $ \gets $ MIN($\frac{numbersNodes(TL_{(i,j,k)})}{2}$, number of nodes of $TL_{(i,j,k)}$ that fit in $cm_{(i,j,l)}$);} \label{divide}

  	\STATE {$Load \gets$ nodes($TL_{(i,j,k)}$, numNodes);}
	 \STATE{\textbf{Send} $Load$ \textbf{to} $T_{(i,j,l)}$; }
     \ELSE
     \STATE{\textbf{Send} $NoLoad$ \textbf{to} $T_{(i,j,l)}$;} 
    \ENDIF
\end{algorithmic} 
\label{ThreadRecvLoadRequest}
\end{algorithm}
\end{minipage}
\end{minipage}

}
\captionof{figure}{Algorithms executed by works threads.}
\end{center}

\fbox{
\def\baselinestretch{1}
\begin{minipage}{0.9\textwidth}
\centering
\begin{minipage}{0.8\textwidth}
\begin{algorithm}[H]
\caption{When $MT_i$ receives $LoadRequest$ from $T_{(i,j,k)}$}
\def\baselinestretch{1}
\begin{algorithmic}[1]
	\IF {(totalIdle = $NT$) and ($x+1 < m$)}
		\STATE {$x++$;}
	    \STATE{\textbf{Send} $LoadRequest$ \textbf{to} $MT_x$;} 
	\ENDIF
	\STATE{totalIdle++;}
\end{algorithmic}
\label{MTRecvLoadRequestT}
\end{algorithm}
\end{minipage}

\begin{minipage}{0.8\textwidth}
\begin{algorithm}[H]
\caption{When $MT_i$ receives $GlobalLoadRequest$ from $MT_x$}
\def\baselinestretch{1}
\begin{algorithmic}[1] 
\STATE {$ML_{i} \gets GlobalLoadRequest$;}
\STATE {numNodes $ \gets \frac{numberNodes(GlobalLoadRequest)}{totalIdle}$;}
\FORALL {($j \gets$ 0 to $j < p$)}
   	\FORALL {($k \gets $ 0 to $k < c$)} 
    		\STATE {$Load \gets$ nodes($ML_i$, numNodes);}  
    		\STATE{\textbf{Send} $Load$ \textbf{to} $T_{(i,j,k)}$;} 
    \ENDFOR
\ENDFOR 
\end{algorithmic}
\label{MTRecvGLoadMT}
\end{algorithm}
\end{minipage}

\begin{minipage}{0.8\textwidth}
\begin{algorithm}[H]
\caption{When $MT_i$ receives $NoLoad$ from $MT_x$}
\def\baselinestretch{1}
\begin{algorithmic}[1]
	\IF {($x+1 < m-1$)}
		\STATE {$x++$;}
		\STATE{\textbf{Send} $LoadRequest$ \textbf{to} $MT_x$;}
	\ELSE 
	\STATE {$terminationDetection()$;}
	\ENDIF
\end{algorithmic}
\label{MTRecvNotLoadMT}
\end{algorithm}
\end{minipage}

\begin{minipage}{0.8\textwidth}
\begin{algorithm}[H]
\caption{When $MT_i$ receives $LoadRequest$ from $MT_x$}
\def\baselinestretch{1}
\begin{algorithmic}[1] 
	\STATE{numLoad $\gets$ 0;}	
	\FORALL {($j \gets$ 0 to $j < p$)} 
		\FORALL {($k \gets $ 0 to $k < c$)}
			\STATE{\textbf{Send} $LoadRequest$ \textbf{to} $T_{(i,j,k)}$;}
		\ENDFOR 
	\ENDFOR 
\end{algorithmic}
\label{MTRecvLoadRequestMT}
\end{algorithm}
\end{minipage}

\begin{minipage}{0.8\textwidth}
\begin{algorithm}[H]
\caption{When $MT_i$ receives $Load$ from $T_{(i,j,k)}$}
\def\baselinestretch{1}
\begin{algorithmic}[1] 
	\STATE{numLoad++;}
	\STATE{$GlobalLoadRequest \gets GlobalLoadRequest + Load$;}
	\IF{(numLoad $ \gets p*c$)}
		\IF{($GlobalLoadRequest \ne \emptyset$)}
			\STATE{\textbf{Send} $GlobalLoadRequest$ \textbf{to} $MT_x$;}
		\ELSE
			\STATE{\textbf{Send} $NoLoad$ \textbf{to} $MT_x$;}
		\ENDIF
	\ENDIF     
\end{algorithmic}
\label{MTSendGLoadMT}
\end{algorithm}
\end{minipage}

\end{minipage}

}
\captionof{figure}{Algorithms executed by manager thread.}

%-----------------------------------------------------

\section{Experimental results}
\label{resultados}

The experiments presented in this section were  executed in  two clusters: Cluster Rio, that was described in Section 3, and   Cluster Oscar, described next.  Each machine of Oscar has two \textit{quad-core} processors   (Intel Xeon 5355 Clovertown). Each \textit{core} has one private L1 cache (64 KB) and  share one  L2 cache (8MB) with another core on the same processor.  All  cores of  a same machine have a uniform access to a 16GB main memory module. Cent OS 5.3 is the operating system  with kernel é 2.6.18.

\subsection{Analyzing Memory Allocation in $PBB_{SPP}$}
\label{memoria}

As seen in the previous section, each node of the $B\&B$ tree can produce other ones, and as a matter of evaluation, both depth and breadth tree traversals were tested in this work.

When breadth traversal was used, the generated nodes, kept in the list $TL_{(i,j,k)}$, occupied more memory than the available space in  L2 cache memory. Although many generated nodes in $TL_{(i,j,k)}$ guarantee that there will be load to be shared with eventually idle cores, it can cause cache access contention. 
In order to certificate that the proposed model  MCM can  be  successfully applied in a real application, we executed the  parallel $B\&B$ several times  varying the   maximum size of $TL_{(i,j,k)}$. Remark that, by following the model, each thread should not use more than the total cache size divided by the number of cores that share it.  In our environment, it means that each one of the two threads  allocated in neighbor cores should use up to 3 MB of L2 cache.
The $PBB_{SPP}$ was executed with the following  $TL_{(i,j,k)}$ maxima sizes: 1, 3, 6  and 8MB.  Tests were performed in one machine of  Cluster Rio. Note that although a breadth traversal procedures is being used, when the size limit of $TL_{(i,j,k)}$ is reached, the algorithm starts the execution of a recursive depth traversal procedure.

Results are presented in the Table \ref{cache}, where columns $|TL_{(i,j,k)}|$, $\#Nodes$, \textit{Wall Clock Time}, \textit{\%CM}, are the maximum size of $TL_{(i,j,k)}$, the number of nodes solved in the corresponding $B\&B$ tree, the wall clock time in seconds of the $PBB_{SPP}$ and the average number of  cache misses for each thread and . Note that these results are averages of ten executions, and in all of the cases the standard deviation was negligible.

The presented wall-clock times  show that  as the $TL_{(i,j,k)}$ size increases, even executing similar number of nodes,  the execution times also increase.  Particularly,  an abrupt  time growing  occurs when  the total $TL_{(i,j,k)}$  size exceeds the L2 cache size, confirming the ability of the proposed model MCM to represent memory contention. Moreover, it can also be observed that cache miss percentage increases with the $TL_{(i,j,k)}$ sizes.

\begin{table*}[htbp]
\centering
\begin{center}
\footnotesize
\caption{Analysis in the number of $B\&B$ tree nodes, the wall clock time end number of caches miss when breath transversal is carried out}
\begin{tabular}{|c|c|r|r|r|}
\hline
\textbf{Instances} & $|TL_{(i,j,k)}|$ & \multicolumn{1}{c|}{\textbf{\# Nodes}} & \multicolumn{1}{c|}{\textbf{Wall clock Time (s)}} & \multicolumn{1}{c|}{\textbf{\%CM}} \\ \hline
I90-400-0.03 & 1MB & 17067 & 10.55 & 28.1777 \\
 & 3MB & 17067 & 11.49 & 29.2282 \\ 
 & 6MB & 17067 & 11.58 & 29.3848 \\ 
 & 9MB & 17067 & 12.09 & 29.4519 \\ \hline
I90-400-0.04 & 1MB & 107205 & 41.09 & 28.9852 \\ 
 & 3MB & 107205 & 44.24 & 30.3375 \\ 
 & 6MB & 107205 & 44.29 & 31.0947 \\ 
 & 9MB & 107205 & 46.67 & 31.3874 \\ \hline
I90-400-0.05 & 1MB & 279272 & 89.55 & 29.1691 \\ 
 & 3MB & 279272 & 94.67 & 30.3543 \\ 
 & 6MB & 279272 & 101.12 & 31.2821 \\ 
 & 9MB & 279272 & 104.65 & 31.3414 \\ \hline
I100-500-0.03 & 1MB & 28641 & 20.50 & 26.3336 \\ 
 & 3MB & 28641 & 22.36 & 27.0692 \\ 
 & 6MB & 28641 & 22.45 & 27.2445 \\ 
 & 9MB & 28641 & 24.04 & 27.3580 \\ \hline
I100-500-0.04 & 1MB & 409252 & 201.60 & 27.7242 \\ 
 & 3MB & 409252 & 213.65 & 28.4698 \\ 
 & 6MB & 409252 & 222.93 & 28.5550 \\ 
 & 9MB & 409252 & 225.28 & 29.3350 \\ \hline
I100-500-0.05 & 1MB & 1999934 & 638.46 & 29.5558 \\ 
 & 3MB & 1999934 & 645.62 & 29.8931 \\ 
 & 6MB & 1999934 & 648.46 & 29.9513 \\ 
 & 9MB & 1999934 & 679.46 & 29.9788 \\ \hline
I110-750-0.03 & 1MB & 20439643 & 15704.29 & 30.7038 \\ 
 & 3MB & 20439643 & 15918.14 & 30.8955 \\ 
 & 6MB & 20439643 & 16789.79 & 31.4379 \\ 
 & 9MB & 20439643 & 30764.23 & 32.3299 \\ \hline
I200-650-0.02-100 & 1MB & 12919402 & 18197.06 & 32.4007 \\ 
 & 3MB & 12919402 & 34126.15 & 32.5090 \\ 
 & 6MB & 12919402 & 35248.22 & 35.0640 \\ 
 & 9MB &  12919402 & 63456.87 & 35.1265 \\ \hline
 
I200-650-0.02-152 & 1MB & 24294476 & 29855.11 & 33.1098 \\ 
 & 3MB & 24294476 & 34644.88 & 33.6970 \\ 
 & 6MB & 24294476 & 113764.58 & 33.1544 \\ 
 & 9MB & 24294476 & 270142.70 &  33.8493\\ \hline
\end{tabular}
\label{cache}
 \end{center}
\end{table*}

\subsection{Evaluating the Load Balance Framework}

In order to evaluate the efficiency of the proposed  Load Balance Framework, $PBB_{SPP}$ was executed both in accordance  with the proposed framework and also  without any load balance procedure.
Tests were executed on two machines of the Oscar cluster running eight threads, one at each core. In the version that no load balance procedure was applied, only the Initial Distribution procedure in Algorithm is executed, and when a thread finishes its nodes, it stays idle until all threads also finish their executions and the application terminates.  In order to evaluate the quality of the load distribution proposed in $PBB_{SPP}$ the following unbalance factor was calculated in accordance with the generated results: $ Un\_Factor=1-\frac{TMed}{TMax}$, \cite{Ma} where $TMed$ is the average of execution times of all the threads and  $TMax$ is the longest execution time among all of them.

Table \ref{semCom}  presents,  for both versions with load balance framework and without load balance,  for each instance, the average of ten executions in seconds (Total Time), the average  of the number of processed nodes in the corresponding $B\&B$ tree (\# Nodes),  the unbalance factor  ($Un\_Factor$),  the coefficient of variation concerning  execution times (CV), and the obtained speedup and efficiencies (E).

 As can be  seen in Table \ref{semCom} executing $PBB_{SPP}$ under the proposed load balance framework doubled the efficiency, even  processing similar number of nodes in most cases. It can also be noted  that the unbalance factor was almost zero for all instances,  indicating the the proposed $PBB_{SPP}$ can really  improve the  application performance.

No results were provided to the instances I110-750-0.04, I110-750-0.05 and I200-600-0.04 since they were executed for more than three days and their execution were halted due to lack of available memory. This is happened because of the initial poor load division.

\begin{table*}[htbp]
\centering
\footnotesize
\caption{ Comparison between the $PBB_{SPP}$  load balance mechanism and a parallel $B\&B$ without load balancing for the same problem}
\begin{tabular}{|l|r|r|r|r|r|r|}
\hline
\textbf{Instances} & \textbf{Total Time (s)} & \textbf{\# Nodes} & \textbf{Un\_Factor} & \textbf{CV} & \textbf{Speedup} & \textbf{E} \\ \hline
\multicolumn{7}{|c|}{\textbf{Without Load Balance}}  \\ \hline
\textbf{I90-400-0.03} & 13.66 & 31481.44 & 0.6370 & 0.36 & 0.68 & 0.04 \\ \hline
\textbf{I90-400-0.04} & 28.31 & 117999.67 & 0.5533 & 0.12 & 0.74 & 0.05 \\ \hline
\textbf{I90-400-0.05} & 43.36 & 232304.80 & 0.7525 & 0.10 & 0.46 & 0.03 \\ \hline
\textbf{I100-500-0.03} & 29.66 & 112652.00 & 0.3200 & 0.24 & 0.67 & 0.04 \\ \hline
\textbf{I100-500-0.04} & 168.34 & 803151.34 & 0.5564 & 0.22 & 0.34 & 0.02 \\ \hline
\textbf{I100-500-0.05} & 723.17 & 1861401.20 & 0.7618 & 0.00 & 0.97 & 0.06 \\ \hline
\textbf{I110-750-0.03} & 6279.81 & 37533812.60 & 0.6317 & 0.29 & 0.37 & 0.02 \\ \hline
\textbf{I110-750-0.04} &  - & - & -  & -  & -  & -   \\ \hline
\textbf{I110-750-0.05} &  - & - & -  & -  & -  & -   \\ \hline
\textbf{I200-650-0.02-100} & 15278.05 & 13032890.60 & 0.8010 & 0.25 & 0.86 & 0.05 \\ \hline
\textbf{I200-650-0.02-152} & 36100.79 & 24354320.17 & 0.8570 & 0.03 & 1.15 & 0.07 \\ \hline
\textbf{I200-600-0.04} &  - & - & -  & -  & -  & -   \\ \hline

\multicolumn{ 7}{|c|}{\textbf{With Load Balance $PBB_{SPP}$}} \\ \hline 
\textbf{I90-400-0.03} & 5.89 & 26028.11 & 0.0161 & 0.20 & 3.43 & 0.21 \\ \hline
\textbf{I90-400-0.04} & 17.39 & 115415.60 & 0.0074 & 0.11 & 2.21 & 0.14 \\ \hline
\textbf{I90-400-0.05} & 37.78 & 280954.67 & 0.0043 & 0.02 & 2.48 & 0.15 \\ \hline
\textbf{I100-500-0.03} & 23.68 & 102316.00 & 0.0075 & 0.18 & 1.88 & 0.12 \\ \hline
\textbf{I100-500-0.04} & 119.14 & 804957.80 & 0.0019 & 0.18 & 4.11 & 0.26 \\ \hline
\textbf{I100-500-0.05} & 267.14 & 2075008.78 & 0.0007 & 0.04 & 2.78 & 0.17 \\ \hline
\textbf{I110-750-0.03} & 5893.70 & 29515686.90 & 0.0000 & 0.22 & 2.90 & 0.18 \\ \hline
\textbf{I110-750-0.04} & 39343.88 & 143389240.00 & 0.0000 & 0.05 & 1.18 & 0.07 \\ \hline
\textbf{I110-750-0.05} & 20018.02 & 106427367.00 & 0.0074 & 0.03 & 1.92 & 0.12 \\ \hline
\textbf{I200-650-0.02-100} & 5690.85 & 13006009.90 & 0.0001 & 0.13 & 3.10 & 0.19 \\ \hline
\textbf{I200-650-0.02-152} & 9786.27 & 24337676.10 & 0.0001 & 0.18 & 3.22 & 0.20 \\ \hline
\textbf{I200-600-0.04} & 30200.41 & 132296456.50 & 0.0000 & 0.01 & 1.92 & 0.12 \\ \hline

\end{tabular}
\label{semCom}
\end{table*}

To  measure the overhead of the proposed $PBB_{SPP}$, distinct phases of the load balance framework was evaluated.  The number of load requests sent  inside a machine and  transmitted to a different machine are shown in Table \ref{overhead}.  As presented in columns, $Local\_Reg$ and  $Global\_Req$, the number of  messages exchanged inside a machine  is much higher than the one  among different machines. Nonetheless, the time of transmitting such messages contribute much less to the total execution time than the messages sent  via network.

\begin{table}[htbp]
\centering
\footnotesize
\caption{Information on the communication}
\begin{tabular}{|l|c|c|c|c|}
\hline
& \multicolumn{2}{|c|}{\textbf{Local}}  & \multicolumn{2}{|c|}{\textbf{Global}}  \\ \hline

\multicolumn{1}{|c|}{\textbf{Instances}} & \textbf{\% Time} & \textbf{\#$Local\_Req$} & \textbf{ \% Time} & \textbf{\# $Global\_Req$} \\ \hline
\textbf{I90-400-0.03} & 3.155 & 50.854 & 12.456 & 5.500 \\ \hline
\textbf{I90-400-0.04} & 1.594 & 125.113 & 8.226 & 10.450\\ \hline
\textbf{I90-400-0.05} & 0.276 & 78.597 & 0.845 & 5.056 \\ \hline
\textbf{I100-500-0.03} & 2.185 & 108.000 & 7.149 & 9.600  \\ \hline
\textbf{I100-500-0.04} & 0.670 & 231.631 & 2.288 & 13.700 \\ \hline
\textbf{I100-500-0.05} & 0.154 & 249.896 & 0.630 & 10.684  \\ \hline
\textbf{I110-750-0.03} & 0.161 & 1381.688 & 0.448 & 36.450  \\ \hline
\textbf{I110-750-0.04} & 0.066 & 1149.050 & 0.131 & 33.200 \\ \hline
\textbf{I110-750-0.05} & 0.053 & 657.464 & 0.139 & 7.699 \\ \hline
\textbf{I200-650-0.02-100} & 0.048 & 647.656 & 0.114 & 19.931 \\ \hline
\textbf{I200-650-0.02-152} & 0.046 & 906.531 & 0.108 & 26.450 \\ \hline
\textbf{I200-600-0.04} & 0.050 & 753.875 & 0.095 & 22.000  \\ \hline
\end{tabular}
\label{overhead}
\end{table}

\subsection{Scalability Experiments}

The last experiment aims to verify the scalability of the $PBB_{SPP}$, by increasing the number of machines available to execute the respective instance. Initially, only two machines were considered in order to measure the messages size exchanged between them. This was carried out to evaluate their impact on the application performance, since as seen in  Section \ref{modelcomphase},  MCM indicated that long messages sent via network might reduce performance. As shown in Table \ref{msgs}, the messages were never longer than 4 Mbytes where \textit{Largest}, indicates the size of the largest message when running the respective application instance, \textit{Smallest}, the size of the smallest message and \textit{Average}, the average amongst all messages size. Secondly, it was also considered four and eight machines, and consequently, more threads were work in parallel. As shown in Tables \ref{4maq8threads} and \ref{8maq8threads}, even with the growing number of messages transmitted via network, performance  was  still improved by $PBB_{SPP}$.

Note that, the messages sizes were never longer than 4MB, therefore priority was given to condition \ref{cond2a} from the Load Balance Model in section \ref{schedModelSec} other than \ref{cond2b}. However, due the amount of $B\&B$ free nodes created, more machines were allocated by $PBB_{SPP}$, upon the saturation of caches of current machines.

\begin{table}[htbp]
\footnotesize
\centering
\caption{Size Messages (KB)}
\begin{tabular}{|l|r|r|r|}
\hline
\multicolumn{1}{|c|}{\textit{Instances}} & \multicolumn{1}{c|}{\textit{Largest}} & \multicolumn{1}{c|}{\textit{Smallest}} & \multicolumn{1}{l|}{\textit{Average}} \\ \hline
\textbf{I90-400-0.03} & 591.73 & 8.50 & 336.28 \\ \hline
\textbf{I90-400-0.04} & 480.80 & 29.10 & 310.38  \\ \hline
\textbf{I90-400-0.05} & 174.81 & 3.32 & 75.38  \\ \hline
\textbf{I100-500-0.03} & 770.92 & 16.48 & 403.37  \\ \hline
\textbf{I100-500-0.04} & 982.40 & 8.28 & 524.22  \\ \hline
\textbf{I100-500-0.05} & 810.58 & 4.43 & 351.16  \\ \hline
\textbf{I110-750-0.03} & 3468.72 & 11.08 & 1903.83  \\ \hline
\textbf{I110-750-0.04} & 3076.16 & 20.02 & 736.14  \\ \hline
\textbf{I110-750-0.05} & 215.04 & 14,76 & 839.86  \\ \hline
\textbf{I200-650-0.02-100} & 1730.42 & 86.31 & 894.08  \\ \hline
\textbf{I200-650-0.02-152} & 1279.84 & 30.08 & 748.10  \\ \hline
\textbf{I200-600-0.04} & 1395.79 & 4.92 & 736.14 \\ \hline
\end{tabular}
\label{msgs}
\end{table}

\begin{table}[htbp]
\centering
\tiny
\caption{$PBB_{SPP}$ execution on four machines}
\begin{tabular}{|l|r|r|r|r|r|r|r|r|}
\hline
Instances & \multicolumn{1}{l|}{Time} & \multicolumn{1}{l|}{\# Nodes} & \multicolumn{1}{l|}{\% Time $Local$} & \multicolumn{1}{l|}{\# $Local\_Req$} & \multicolumn{1}{l|}{\% Time $Global$} & \multicolumn{1}{l|}{\# $Global\_Req$} & \multicolumn{1}{l|}{$Un\_Factor$} & \multicolumn{1}{l|}{Speedup} \\ \hline
\textbf{I90-400-0.03} & 4.49 & 25383 & 5.2656 & 50.1250 & 35.4018 & 11.2500 & 0.0339 & 4.505 \\ \hline
\textbf{I90-400-0.04} & 10.44 & 115431 & 1.9193 & 115.3438 & 15.3741 & 18.0000 & 0.0084 & 3.687 \\ \hline
\textbf{I90-400-0.05} & 19.66 & 282458 & 1.0993 & 162.5000 & 7.8492 & 19.0000 & 0.0073 & 4.765 \\ \hline
\textbf{I100-500-0.03} & 11.83 & 90535 & 3.1212 & 90.7188 & 18.2631 & 15.5000 & 0.0131 & 3.757\\ \hline
\textbf{I100-500-0.04} & 94.38 & 1166225 & 0.8258 & 291.9375 & 5.6362 & 26.5000 & 0.0022 & 5.184 \\ \hline
\textbf{I100-500-0.05} & 148.38 & 2247700 & 0.9256 & 401.2188 & 2.4457 & 24.5000 & 0.0041 & 49.990 \\ \hline
\textbf{I110-750-0.03} & 2345.01 & 20947396 & 0.3405 & 1492.2813 & 1.2305 & 66.0000 & 0.0000 & 7.291 \\ \hline
\textbf{I110-750-0.04} & 11338.25 & 127309716 & 0.1154 & 1368.5000 & 0.3227 & 47.0000 & 0.0000 & 4.106 \\ \hline
\textbf{I110-750-0.05} & 7349.01 & 111828773 & 0.0985 & 932.8438 & 0.4027 & 29.2500 & 0.0001 & 5.217 \\ \hline
\textbf{I200-650-0.02-100} & 2583.50 & 12962936 & 0.0792 & 675.1875 & 0.2889 & 29.7742 & 0.0001 & 6.839 \\ \hline
\textbf{I200-650-0.02-152} & 4692.99 & 24327659 & 0.0623 & 931.4688 & 0.2137 & 37.2500 & 0.0000 & 6.706\\ \hline
\textbf{I200-600-0.04} & 9834.11 & 130502067 & 0.1349 & 1420.6875 & 0.3329 & 44.5000 & 0.0001 & 5.907 \\ \hline
\end{tabular}
\label{4maq8threads}
\end{table}

\begin{table}[htbp]
\centering
\tiny
\caption{$PBB_{SPP}$ execution on eight machines}
\begin{tabular}{|l|r|r|r|r|r|r|r|r|}
\hline
Instances & \multicolumn{1}{l|}{Time} & \multicolumn{1}{l|}{\# Nodes} & \multicolumn{1}{l|}{\% Time $Local$} & \multicolumn{1}{l|}{\# $Local\_Req$} & \multicolumn{1}{l|}{\% Time $Global$} & \multicolumn{1}{l|}{\# $Global\_Req$} & \multicolumn{1}{l|}{$Un\_Factor$} & \multicolumn{1}{l|}{Speedup} \\ \hline
\textbf{I90-400-0.03} & 3.70 & 39845 & 0.204 & 56.609 & 1.726 & 15.938 & 0.035 & 5.463 \\ \hline
\textbf{I90-400-0.04} & 7.34 & 119475 & 0.249 & 109.953 & 2.613 & 18.000 & 0.018 & 5.242 \\ \hline
\textbf{I90-400-0.05} & 13.46 & 289197 & 0.242 & 120.570 & 2.977 & 16.375 & 0.016 & 6.962 \\ \hline
\textbf{I100-500-0.03} & 6.30 & 71820 & 0.298 & 72.016 & 2.030 & 16.125 & 0.023 & 7.052 \\ \hline
\textbf{I100-500-0.04} & 38.20 & 858877 & 0.589 & 203.945 & 3.208 & 24.813 & 0.004 & 12.807 \\ \hline
\textbf{I100-500-0.05} & 76.49 & 2186299 & 0.491 & 292.836 & 4.059 & 24.438 & 0.002 & 9.698 \\ \hline
\textbf{I110-750-0.03} & 5133.82 & 95623639 & 7.539 & 1325.859 & 23.942 & 69.375 & 0.000 & 3.331 \\ \hline
\textbf{I110-750-0.04} & 10256.64 & 248729987 & 11.959 & 1244.086 & 34.212 & 50.250 & 0.000 & 4.539 \\ \hline
\textbf{I110-750-0.05} & 3239.82 & 112034575 & 3.544 & 861.578 & 13.866 & 38.250 & 0.000 & 11.835 \\ \hline
\textbf{I200-650-0.02-100} & 1476.05 & 12948620 & 1.930 & 692.391 & 9.250 & 39.250 & 0.000 & 11.969 \\ \hline
\textbf{I200-650-0.02-152} & 2240.29 & 24332087 & 2.642 & 819.422 & 13.442 & 44.625 & 0.000 & 14.048 \\ \hline
\textbf{I200-600-0.04} & 4255.59 & 131120162 & 5.959 & 1273.219 & 21.805 & 50.625 & 0.000 & 13.651 \\ \hline

\end{tabular}
\label{8maq8threads}
\end{table}

\section{Conclusions and Future work}
\label{}

This paper proposes  the MCM model that represents the most relevant characteristics of a multicore cluster, based on the results of exhaustive experiments of a synthetic application.  In order to validate the model, it was used in the design  and development of  a Parallel Branch-and-Bound for the Set Partitioning Problem .Under the MCM, a load balance framework  for solving this problem  prevents  that memory contention directly affects the performance, scheduling the nodes of the $B\&B$ tree accordingly to the available amount of the cache memory. It was shown that the bottlenecks are avoided since the execution times  improved considerably. Further analyzes will be conducted for the model on other classes of application. The actual application used is considered to be dynamic, and therefore, other applications with different characteristics will be considered in future work in order to show the efficiency of the model.

\bibliographystyle{wileyj}
\bibliography{journal}

\begin{thebibliography}{10}
\providecommand{\url}[1]{\texttt{#1}}
\providecommand{\urlprefix}{URL }
\expandafter\ifx\csname urlstyle\endcsname\relax
  \providecommand{\doi}[1]{doi:\discretionary{}{}{}#1}\else
  \providecommand{\doi}{doi:\discretionary{}{}{}\begingroup
  \urlstyle{rm}\Url}\fi

\bibitem{Savage}
Savage JE, Zubair M. A unified model for multicore architectures.
  \emph{Proceedings of the 1st international forum on Next-generation
  multicore/manycore technologies}, IFMT '08, ACM: New York, NY, USA, 2008;
  9:1--9:12.

\bibitem{Song2011}
Tang L, Mars J, Soffa ML. Contentiousness vs. sensitivity: improving contention
  aware runtime systems on multicore architectures. \emph{Proceedings of the
  1st International Workshop on Adaptive Self-Tuning Computing Systems for the
  Exaflop Era}, EXADAPT '11, ACM: New York, NY, USA, 2011; 12--21.

\bibitem{Sadaf}
Alam SR, Barrett RF, Kuehn JA, Roth PC, Vetter JS. Characterization of
  scientific workloads on systems with multi-core processors. \emph{IEEE
  International Symposium on Workload Characterization}, IISWC, IEEE, 2006;
  225--236.

\bibitem{Song}
Song F, YarKhan A, Dongarra J. Dynamic task scheduling for linear algebra
  algorithms on distributed-memory multicore systems. \emph{International
  Conference for High Performance Computing, Networking Storage and Analysis,},
  2009.

\bibitem{Mars}
Mars J, Tang L, Soffa ML. Directly characterizing cross core interference
  through contention synthesis. \emph{Proceedings of the 6th International
  Conference on High Performance and Embedded Architectures and Compilers},
  HiPEAC '11, ACM: New York, NY, USA, 2011; 167--176.

\bibitem{Rashid}
Rashid H, Novoa C, Qasem A. An evaluation of parallel knapsack algorithms on
  multicore architectures. \emph{CSC'10}, 2010; 230--235.

\bibitem{Wyllie}
Fortune S, Wyllie J. Parallelism in random access machine. \emph{10th ACM
  Symposium on Theory of Computation (STOC)}, New York, USA, 1978; 114--118.
  \urlprefix\url{http://budiu.info/work/ipdps11.pdf}.

\bibitem{JaJa}
Jajá J. \emph{An introduction to parallel algorithms}. Addison Wesley Longman
  Publishing Co., Inc.: Redwood City, CA, USA, 1992.

\bibitem{Cole}
Cole R, Zajicek O. The \textit{APRAM} : incorporating asynchrony into the
  \textit{PRAM} model. \emph{Proceedings of the first annual ACM symposium on
  Parallel algorithms and architectures}, SPAA '89, ACM: New York, NY, USA,
  1989; 169--178.

\bibitem{Gibbons}
Gibbons PB. A more practical \textit{PRAM} model. \emph{Proceedings of the
  first annual ACM symposium on Parallel algorithms and architectures}, SPAA
  '89, ACM: New York, NY, USA, 1989; 158--168.

\bibitem{Gibbons2}
Gibbons PB, Matias Y, Ramachandran V. Can shared-memory model serve as a
  bridging model for parallel computation? \emph{Proceedings of the ninth
  annual ACM symposium on Parallel algorithms and architectures}, SPAA '97,
  ACM: New York, NY, USA, 1997; 72--83.

\bibitem{Gibbons3}
Gibbons PB, Matias Y, Ramachandran V. The \textit{QRQW PRAM}: accounting for
  contention in parallel algorithms. \emph{Proceedings of the fifth annual
  ACM-SIAM symposium on Discrete algorithms}, SODA '94, Society for Industrial
  and Applied Mathematics: Philadelphia, PA, USA, 1994; 638--648.

\bibitem{Maggs}
Maggs BM, Matheson LR, Tarjan RE. Models of parallel computation: A survey and
  synthesis 1995.

\bibitem{Vijaya}
Ramachandran V. \textit{QSM}: A general purpose shared-memory model for
  parallel computation. \emph{Foundations of Software Technology and
  Theoretical Computer Science}, 1997; 1--5.

\bibitem{Aggarwal}
Aggarwal A, Alpern B, Chandra A, Snir M. A model for hierarchical memory.
  \emph{Proceedings of the nineteenth annual ACM symposium on Theory of
  computing}, STOC '87, ACM: New York, NY, USA, 1987; 305--314.

\bibitem{Aggarwal2}
Aggarwal A, Chandra AK, Snir M. Hierarchical memory with block transfer.
  \emph{Proceedings of the 28th Annual Symposium on Foundations of Computer
  Science}, SFCS '87, IEEE Computer Society: Washington, DC, USA, 1987;
  204--216.

\bibitem{Ben}
Juurlink B, Juurlink BHH, Wijshoff HAG. The parallel hierarchical memory model.
  \emph{In Proc. Scandinavian Workshop on Algorithms Theory, LNCS 824},
  Springer-Verlag, 1994; 240--251.

\bibitem{Bowen1}
Alpern B, Carter L, Ferrante J. Modeling parallel computers as memory
  hierarchies. \emph{In Proc. Programming Models for Massively Parallel
  Computers}, IEEE Computer Society Press, 1993; 116--123.

\bibitem{Bowen}
Alpern B, Carter L, Feig E, Selker T. The uniform memory hierarchy model of
  computation. \emph{Algorithmica}  1994; \textbf{12}:72--109.

\bibitem{Papadimitriou}
Papadimitriou C, Yannakakis M. Towards an architecture-independent analysis of
  parallel algorithms. \emph{STOC '88: Proceedings of the twentieth annual ACM
  symposium on Theory of computing}, ACM: New York, NY, USA, 1988; 510--513.

\bibitem{Sena}
Sena A. Um modelo alternativo para execução eficiente de aplicações paralelas
  \textit{MPI} nas grades computacionais. Ph{D} {T}hesis, Universidade Federal
  Fluminense 2008.

\bibitem{Valiant}
Valiant LG. A bridging model for parallel computation. \emph{Commun. ACM}
  1990; \textbf{33}(8):103--111.

\bibitem{Williams}
Williams TL, Parsons RJ. The heterogeneous bulk synchronous parallel model.
  \emph{Proceedings of the 15 IPDPS 2000 Workshops on Parallel and Distributed
  Processing}, IPDPS '00, Springer-Verlag: London, UK, UK, 2000; 102--108.

\bibitem{Culler}
Culler DE, Karp RM, Patterson D, Sahay A, Santos EE, Schauser KE, Subramonian
  R, von Eicken T. \textit{LogP}: a practical model of parallel computation.
  \emph{Commun. ACM}  November 1996; \textbf{39}:78--85.

\bibitem{Alexandrov}
Alexandrov A, Ionescu MF, Schauser KE, Scheiman C. {LogGP}: Incorporating long
  messages into the {LogP} model - one step closer towards a realistic model
  for parallel computation 1995.

\bibitem{Ino}
Ino F, Fujimoto N, Hagihara K. \textit{LogGPS}: a parallel computational model
  for synchronization analysis. \emph{SIGPLAN Not.}  June 2001;
  \textbf{36}:133--142.

\bibitem{Frank}
Frank M, Agarwal A, Vernon MK. \textit{LoPC}: Modeling contention in parallel
  algorithms. \emph{PPOPP}, 1997; 276--287.

\bibitem{Cameron}
Cameron KW, Ge R, Sun XH. $\log_{\rm n}{\rm \textit{p}}$ and $\log_{3}{\rm p}$:
  Accurate analytical models of point-to-point communication in distributed
  systems. \emph{IEEE Transactions on Computers}  2007; \textbf{56}:314--327,
  \doi{10.1109/TC.2007.38}.

\bibitem{Bilardi}
Bilardi G, Herley KT, Pietracaprina A, Pucci G, Spirakis PG. Bsp vs logp.
  \emph{SPAA}, 1996; 25--32.

\bibitem{Ramachandran}
Ramachandran V, Grayson B, Dahlin M. Emulations between \textit{QSM, BSP and
  LogP}: a framework for general-purpose parallel algorithm design. \emph{J.
  Parallel Distrib. Comput.}  2003; \textbf{63}(12):1175--1192.

\bibitem{Tam}
Tam AT, Wang CL. Realistic communication model for parallel computing on
  cluster. \emph{1st IEEE Computer Society International Workshop on Cluster
  Computing}, 1999.

\bibitem{Mendes}
de~Amorim~Mendes H. {HlogP} : Um modelo de escalonamento para execução de
  aplicações \textit{MPI} em grades computacionais. Master's {T}hesis,
  Universidade Federal Fluminense 2004.

\bibitem{Badia}
Badia RM, Perez JM, Ayguade E, Labarta J. Impact of the memory hierarchy on
  shared memory architectures in multicore programming models.
  \emph{Proceedings of the 2009 17th Euromicro International Conference on
  Parallel, Distributed and Network-based Processing}, IEEE Computer Society:
  Washington, DC, USA, 2009; 437--445.

\bibitem{Chai}
Chai L, Gao Q, Panda DK. Understanding the impact of multi-core architecture in
  cluster computing: A case study with intel dual-core system.
  \emph{Proceedings of the Seventh IEEE International Symposium on Cluster
  Computing and the Grid}, CCGRID '07, IEEE Computer Society: Washington, DC,
  USA, 2007; 471--478.

\bibitem{Savage2}
Savage JE, Zubair M. Evaluating multicore algorithms on the unified memory
  model. \emph{Scientific Programming - Software Development for Multi-core
  Computing Systems}  December 2009; \textbf{17}:295--308.

\bibitem{Tu}
Tu B, Fan J, Zhan J, Zhao X. Accurate analytical models for message passing on
  multi-core clusters. \emph{Proceedings of the 2009 17th Euromicro
  International Conference on Parallel, Distributed and Network-based
  Processing}, IEEE Computer Society: Washington, DC, USA, 2009; 133--139.

\bibitem{Ortega}
Mercier G, Clet-Ortega J. Towards an efficient process placement policy for
  {MPI} applications in multicore environments. \emph{EuroPVM/MPI},
  \emph{Lecture Notes in Computer Science}, vol. 5759, Springer: Espoo,
  Finland, 2009; 104--115. \urlprefix\url{http://hal.inria.fr/inria-00392581}.

\bibitem{Song2}
Song F, Moore S, Dongarra J. Analytical modeling for affinity-based thread
  scheduling on multicore plataforms. \emph{Symposim onPrinciples and Parctice
  of Parallel Programming}, 2009.

\bibitem{prasanna_grupos}
Xia Y, Prasanna VK, Li J. Hierarchical scheduling of dag structured
  computations on manycore processors with dynamic thread grouping.
  \emph{Proceedings of the 15th international conference on Job scheduling
  strategies for parallel processing}, JSSPP'10, Springer-Verlag: Berlin,
  Heidelberg, 2010; 154--174.

\bibitem{servet}
González-Domínguez J, Taboada GL, Fraguela BB, Martín MJ, Touriño J. Servet: A
  benchmark suite for autotuning on multicore clusters. \emph{IPDPS'10}, 2010;
  1--9.

\bibitem{Saavedra}
Smith AJ, Saavedra RH. Measuring cache and tlb performance and their effect on
  benchmark runtimes. \emph{IEEE Trans. Comput.}  Oct 1995;
  \textbf{44}(10):1223--1235.

\bibitem{papi}
{Innovating Computing Laboratory, University of Tennessee}. Performance
  application programming interface 2004. Http://icl.cs.utk.edu/papi/.

\bibitem{Dongarra}
Dongarra J, Moore S, Mucci P, Seymour K, You H. Accurate cache and tlb
  characterization using hardware counters. \emph{International Conference on
  Computational Science}, Krakow, Poland, 2004.

\bibitem{Bauer}
Barreto L, Bauer M. Parallel branch and bound algorithm - a comparison between
  serial, openmp and mpi implementations. \emph{Journal of Physics: Conference
  Series}  2010; \textbf{256}(1):012\,018.

\bibitem{Bob}
Djerrah A, Cun BL, Cung VD, Roucairol C. Bob++: Framework for solving
  optimization problems with branch-and-bound methods. \emph{HPDC}, 2006;
  369--370.

\bibitem{Cun2011}
Galea F, Cun BL. A parallel exact solver for the three-index quadratic
  assignment problem. \emph{IPDPS Workshops}, 2011; 1940--1949.

\bibitem{Lucia}
de~A~Drummond LM, Uchoa E, Gon\c{c}alves AD, Silva JMN, Santos MCP, de~Castro
  MCS. A grid-enabled distributed branch-and-bound algorithm with application
  on the steiner problem in graphs. \emph{Parallel Computing}  2006;
  \textbf{32}(9):629--642.

\bibitem{SYMPHONY}
Ralphs TK, G{\"u}zelsoy M, Mahajan A. {SYMPHONY} version 5.3 user's manual.
  \emph{Technical {R}eport}, COR@L Laboratory, Lehigh University 2011.
  \urlprefix\url{http://www.coin-or.org/SYMPHONY/doc/SYMPHONY-5.3.4-Manual.pdf%
}.

\bibitem{Estrada}
Sanjuan-Estrada JF, Casado LG, García I. Adaptive parallel interval branch and
  bound algorithms based on their performance for multicore architectures.
  \emph{The Journal of Supercomputing}  2011; :376--384.

\bibitem{Kim}
Park S, Kim T, Park J, Kim J, Im H. Parallel skyline computation on multicore
  architectures. \emph{Proceedings of the 2009 IEEE International Conference on
  Data Engineering}, ICDE '09, IEEE Computer Society: Washington, DC, USA,
  2009; 760--771.

\bibitem{PICO}
Eckstein J, Phillips CA, Hart WE. Pico: An object-oriented framework for
  parallel branch and bound 2001.

\bibitem{PUBB}
Shinano Y, Higaki M, Hirabayashi R. A generalized utility for parallel branch
  and bound algorithms. \emph{Proceedings of the 7th IEEE Symposium on Parallel
  and Distributeed Processing}, SPDP '95, IEEE Computer Society: Washington,
  DC, USA, 1995; 392--.

\bibitem{Werneck}
Budiu M, Delling D, Werneck R. {DryadOpt}: Branch-and-bound on distributed
  data-parallel execution engines. \emph{IEEE International Parallel and
  Distributed Processing Symposium (IPDPS)}, Anchorage, AK, 2011.
  \urlprefix\url{http://budiu.info/work/ipdps11.pdf}.

\bibitem{Hochbaum}
Hochbaum DS. \emph{Approximation algorithms for NP-hard problems}. PWS
  Publishing Co.: Boston, MA, USA, 1997.

\bibitem{dual}
Boschetti MA, Mingozzi A, Ricciardelli S. A dual ascent procedure for the set
  partitioning problem. \emph{Discret. Optim.}  Nov 2008;
  \textbf{5}(4):735--747.

\bibitem{Ma}
Ma KL. Parallel volume ray-casting for unstructured-grid data on
  distributed-memory architectures. \emph{Proceedings of the IEEE symposium on
  Parallel rendering}, PRS '95, ACM: New York, NY, USA, 1995; 23--30.

\end{thebibliography}

\end{document}